\documentstyle[floats,aps, prd,epsfig]{revtex}
\begin{document}

\twocolumn[\hsize\textwidth\columnwidth\hsize\csname
@twocolumnfalse\endcsname
%

\title{Locating Boosted Kerr and Schwarzschild Apparent Horizons.}
\author{Mijan F. Huq${}^{(1)}$, 
Matthew W. Choptuik${}^{(2)}$ 
and Richard A. Matzner${}^{(3)}$} 
\address{${}^{(1)}$Department of Astronomy \&
Astrophysics and Center for Gravitational Physics \& Geometry, \\
The Pennsylvania State University, University Park,
PA 16803 \\
${}^{(2)}$ Department of Physics and Astronomy, \\
 University of British Columbia,
Vancouver BC,   V6T 1Z1,   Canada,\\
${}^{(3)}$Center for Relativity, \\
The University of Texas at Austin, Austin, TX
78712-1081} 
\maketitle 
\begin{abstract} 
We describe a finite-difference method for locating apparent horizons and 
illustrate its capabilities on boosted Kerr and Schwarzschild black holes. 
Our model spacetime is given by the Kerr-Schild metric. We apply a Lorentz 
boost to this spacetime metric and then carry out a 3+1 decomposition.  The 
result is a slicing of Kerr/Schwarzschild in which the black hole is 
propagated and Lorentz contracted. We show that our method can locate 
distorted apparent horizons efficiently and accurately.
\end{abstract}

\vskip2pc]
\def\R4{{}^{(4)}R}
\def\g4{{}^{(4)}g}
\def\half{\frac{1}{2}}
\def\lie{\mbox{\pounds}}
\newcommand{\lp}{\left[}
\newcommand{\rp}{\right]}
\newcommand{\HALF}{ \frac{1}{2} }
\newcommand{\SIXTH}{ \frac{1}{6} }
\newcommand{\lc}{\bigl(}
\newcommand{\rc}{\bigl)}
\newcommand{\Lc}{\Bigl(}
\newcommand{\Rc}{\Bigl)}
\newcommand{\lC}{\biggl(}
\newcommand{\rC}{\biggl)}
\newcommand{\LC}{\biggl(}
\newcommand{\RC}{\biggl)}
\newcommand{\lb}{\left(}
\newcommand{\rb}{\right)}

\section{Introduction}

Apparent horizon locators play an integral role in the application of
black hole excision techniques in the computational evolution of black
hole spacetimes. Excision techniques delete the regions of spacetime
that contain the curvature singularity from the computational domain.
Assuming cosmic censorship, these curvature singularities are
expected to be contained within an event horizon. The event horizon is
a causal boundary whose interior does not causally affect the exterior
spacetime; as a result it is possible to excise a region within the
event horizon, thereby excising the black hole's curvature singularity.

In our approach to computationally solving the Einstein field
equations we focus on the use of Cauchy techniques, in which a 3+1
splitting of spacetime into a foliation of spacelike hypersurfaces,
$\Sigma$, parametrized in time, is the basis for an evolution in time.
The result of this splitting is a system of elliptic and hyperbolic
partial differential equations in the 3-metric, $\gamma_{ij}$, and
extrinsic curvature, $K_{ij}$. These are the four constraint equations
and 12 first-order-in-time evolution equations. The Cauchy approach
starts with an initial spacelike slice with $\gamma_{ij}$ and $K_{ij}$
set by solving an initial value problem (the elliptic constraint
equations).  One then uses the evolution equations to evolve to the
next spacelike slice obtaining $\gamma_{ij}$ and $K_{ij}$ at the next
time (See York\cite{York.frontiers} for a detailed discussion).

In the evolution of black hole spacetimes in this manner we do not have
a complete history of the entire spacetime and hence do not have a
knowledge of the location of the event horizon. Since the event horizon
is a global object that depends on geometric information for all
time (or at least until the black hole becomes quiescent) we cannot use
it to determine an inner excision boundary in our Cauchy evolution.
There is, however, an alternative, and that is to use the apparent
horizon surface which is a local object, locatable (if it exists) with
$\gamma_{ij}$ and $K_{ij}$ at one time.  The apparent horizon is the
outermost marginally trapped surface. It is a closed spacelike
2-surface whose future-directed outgoing null normals have
zero divergence\cite{HawkingandEllis}. The apparent horizon is slicing
dependent and may not necessarily exist even though an event horizon
does.  An example of this is given by Wald and Iyer\cite{WaldandIyer}
through nonspherically-symmetric slicings 
for the Schwarzschild spacetime. 
Provided a {\it non-pathological} slicing is chosen the apparent
horizon or any trapped surface within it may be used for excising the
black hole singularity. 
These surfaces define a local causal structure that
distinguishes instantaneously escaping null rays from those that are
certain to collapse. This distinction makes their treatment very
amenable to computational black hole excision techniques. Since these
surfaces can be determined with geometric information at one instant of
time, they are used in practice as an inner boundary in Cauchy evolutions.  
With this purpose in mind, we
developed a 3D apparent horizon locator that utilizes $\gamma_{ij}$ and
$K_{ij}$ on a given spacelike slice of spacetime and locates an
apparent horizon. Once the apparent horizons are located, a region
contained within the apparent horizon is excised. Thus the method is
really a trapped-surface excision.

There has been a variety of work done on apparent horizon location in
spherical symmetry, axisymmetry and 3D. We focus solely on the 3d
locators. These can be classified into those that
use finite difference methods, and those that utilize pseudo-spectral
schemes.  Further, one can classify each of these finders in terms
of those that use flow methods versus those that directly solve the
apparent horizon equation either via a minimization scheme or Newton's
method for root finding.  

One of the first published 3d apparent horizon locators was developed by
Nakamura, Kojima and Oohara\cite{nakamura.1}. Their method expands the 
{\it apparent horizon shape function}, $r=\rho(\theta,\phi)$ in spherical
harmonics to some maximum $l=l_{max}$:
\begin{eqnarray}
\rho(\theta,\phi) & = & \sum_{l=0}^{l_{max}} \sum^{l}_{m=-l} a_{lm}
Y_{lm}(\theta,\phi).
\end{eqnarray}

With this expansion Nakamura et al. evaluate the apparent horizon equation 
and solve
for the coefficients $a_{lm}$ via a ``direct" functional iteration scheme.
Kemball and Bishop\cite{bishop.1} reimplemented this approach and made 
modifications that led to improved convergence and stability behaviour.
Anninos et al.\cite{anninos.1} and Baumgarte et al.\cite{baumgarte.1} 
implement similar methods that involve an expansion of $\rho(\theta,\phi)$;
the primary differences being that they expand in terms of symmetric 
trace-free tensors and use Powell's method for minimization of the square of
the apparent horizon equation, Eq.(2) below (which is related to the 
expansion of the outgoing null normals). 

Thornburg\cite{thornburg.1} gives a very good treatise on the use of finite
differencing to solve the apparent horizon equation using spherical 
coordinates $(r,\theta,\phi)$ via Newton's method. He discusses in general
how algebraic Jacobians may be applied in a full 3D context. His 
implementation for horizon finding however is axisymmetric; his full 
3-d finder suffers from problems with the z-axis ($\theta=0,\pi$). 
Our method for finding horizons uses closely related concepts
except that we finite difference in cartesian coordinates, eliminating
any potential z-axis problems. 

Another class of apparent horizon locators casts the elliptic apparent
horizon equation into a parabolic one as suggested by Tod\cite{Tod},
who suggested the use of flow methods in locating apparent horizons.
Bernstein\cite{bernstein} implemented Tod's algorithm in axisymmetry
using finite differences, but encountered problems with differencing on
a sphere in spherical coordinates in the general case.

The advantage to flow-methods is that one can start with an arbitrary
initial guess and flow towards the apparent horizon(s). In some
implementations it is possible to find multiple apparent horizon
surfaces starting from a single initial guess surface (i.e., there is a
topology change in the course of location of the apparent horizon).
Pasch\cite{pasch} uses a level-set method to locate multiple
apparent horizons in 3d. He demonstrates his method utilizing
time-symmetric conformally flat initial data for multiple black holes.
An hybrid flow/level-set-like method utilizing our approach to
evaluating the outgoing null expansions via Cartesian finite
differences has been implemented by Shoemaker et al.\cite{shoemaker}.
That method flows towards the apparent horizon(s) from an arbitrary
initial guess allowing for topology changes.
Gundlach\cite{Gundlach} has implemented a ``fast flow" method for finding
apparent horizons. 

In the sections that follow we give a brief discussion of the algorithm used
and relegate the details to the appendices. The model
spacetime in which all of the results are presented is discussed in section
III. In section IV we discuss tests of the algorithm and demonstrate that
for distorted apparent horizons for boosted Kerr black holes
the algorithm fairs well. 

\section{Boundary Value Problem Approach}

On a particular 3D spacelike hypersurface, $\Sigma$, from our foliation of 
spacetime we are given the 3-metric, $\gamma_{ij}$ and the
extrinsic curvature, $K_{ij}$. Let ${\cal S}$ be a closed 2-surface in 
$\Sigma$. At any point $p$ on ${\cal S}$ we can define a spacelike normal,
$s^a$, to ${\cal S}$, and a time-like normal, $n^a$, to $\Sigma$. From these
we can construct the outgoing null normal, $k^a$, at $p$. If the divergence
$\nabla_{a} k^{a}$($\nabla_a$ is the covariant derivative compatible with
the spacetime metric, $g_{ab}$) is zero {\it everywhere} on ${\cal S}$, then
${\cal S}$ is a {\it marginally trapped surface} (MTS). 
The apparent horizon is the outermost such MTS. The expansion of the outgoing
null normals, $\nabla_a k^a=0$, can be rewritten as an equation entirely in
terms of quantities in $\Sigma$\cite{York.frontiers}:
\begin{eqnarray}
D_i s^i + K_{ij} s^i s^j - K &=& 0. \label{aheqn.1}
\end{eqnarray}
Here $D_i$ is the covariant derivative compatible with the 3-metric 
$\gamma_{ij}$ and $K$ is the trace of the extrinsic curvature.

The apparent horizon equation is an elliptic partial differential equation on 
${\cal S}$ (for a function of coordinates on ${\cal S}$). This can be made
apparent by noting that an MTS is a closed 2-surface; spherical coordinates
are a natural set of coordinates for ${\cal S}$. The location of ${\cal S}$
can then be written as the radial distance from the origin of
the coordinate system, $r=\rho(\theta,\phi)$. In general one can generate 
a foliation of such closed spacelike 2-surfaces based on the distance from
the MTS. This is given by
\begin{eqnarray}
\varphi &=& r - \rho(\theta,\phi), \label{level.1}
\end{eqnarray}
where the $\varphi = 0$ level surface is the MTS. From $\varphi$ we 
define a spacelike vector field, the normal
\begin{eqnarray}
s^i & = & \gamma^{ij} \partial_j \varphi / \sqrt{\gamma^{kl} \partial_k \varphi \partial_l \varphi }, \label{ahnormal.1}
\end{eqnarray}
at every point on these level surfaces. Substituting this into 
Eq.(\ref{aheqn.1}) results in a second order elliptic partial differential 
equation on ${\cal S}$,
\begin{eqnarray} 
F[\rho] & = & \gamma^{ab}\partial_a \partial_b \varphi + \gamma^{ab}_{~~,a} 
\partial_b \varphi - \frac{1}{2}\omega^{-1} \gamma^{ab}\gamma^{cd}_{a}
\partial_b \varphi \partial_c \varphi \partial_d \varphi \nonumber \\
& & - \omega^{-1} \gamma^{ab}\gamma^{cd} \partial_b \varphi \partial_a 
\partial_c \varphi \partial_d \varphi + \Gamma^{a}_{ab} \gamma^{bc} 
\partial_c \varphi \nonumber \\
& & + \omega^{-\frac{1}{2}} K_{ab} \gamma^{ac}\gamma^{bd}\partial_c \varphi 
\partial_d \varphi - \omega^{\frac{1}{2}} K = 0,
\label{eq:apphor2.pp1}
\end{eqnarray}
where $\omega = \gamma^{cd} \partial_c \varphi \partial_d \varphi$ and
$\Gamma^a_{~bc}$ is the connection coefficient associated with the 3-metric
$\gamma_{ab}$. 

Our approach involves casting Eq.(\ref{eq:apphor2.pp1}) as a boundary
value problem on ${\cal S}$. As stated, points on ${\cal S}$ are
parametrized in spherical coordinates $(\theta\in [0,\pi] , \phi \in
[0,2\pi))$. ${\cal S}$ is discretized into an uniform mesh, $\hat{\cal
S}$, of $N_\theta \times N_\phi$ points where $N_\theta =  N_\phi =
N_s$.  The domain on $\hat{\cal S}$ is $(0 \le \theta \le \pi ; 0 \le
\phi < 2\pi)$ where at the poles, $\theta =0, \pi$ all $N_{\phi}$
points are identified as one. The $\phi=2\pi$ branch cut is identified
with the $\phi=0$ line.  The boundary conditions simply are periodicity
at $\phi=2\pi$ and $\phi$ identification at $\theta=0,\pi$.  
These boundary conditions are key to avoiding the coordinate singularities
at the poles in combination with using {\it Cartesian} coordinates to 
discretize partial derivatives on $\hat{\cal S}$. We treat $\varphi$ as a 
function of Cartesian coordinates $x,y,z$ and center on each mesh point of
$\hat{\cal S}$ a 3-d Cartesian difference stencil of 27 points. Using
the form Eq.(\ref{level.1}) we interpolate values of $\varphi(x,y,z)$
onto each of the 26 stencil points surrounding each $\hat{\cal S}$
stencil point. (See the appendix for more details.) Using this
difference stencil we can evaluate first, second and mixed derivatives
of $\varphi(x,y,z)$ as required by the discretized version of
Eq.(\ref{eq:apphor2.pp1}). At every point on $\hat{\cal S}$ we 
then construct  the residual $\hat{F}[\hat{\rho}]$ on $\hat{\cal S}$
(Note that the discrete version of a continuum quantity, $T$, is
denoted by, $\hat{T}$).

The problem at hand is to solve for a $\rho$ that yields $F[\rho]=0$. Since
$F[\rho]$ is a nonlinear operator (as shown in Eq.(\ref{eq:apphor2.pp1})),
we use Newton's method to solve for $\rho$. Given an initial guess 
surface,
$\rho=\rho_0$, we wish to find a $\delta \rho$  (the change in the surface)
that leads to $F[\rho_0 + \delta \rho] = 0$ or, to lowest order, 
\begin{equation}
F[\rho_0 + \delta \rho] = F[\rho_0] + 
\left.\frac{\partial F[\rho]}{\partial \rho}\right|_{\rho=\rho_0} 
\delta \rho + O(\delta \rho^2) = 0.
\label{eqn:taylor.1}
\end{equation}
The Jacobian of $F[\rho]$ is defined to be 
\begin{equation}
J \equiv \frac{\partial F}{\partial \rho}.
\end{equation}
In the 
discretized case, $\hat{J}$ is an $N\times N$ matrix, where $N$ is the total
number of points used in the discretization.
To obtain a $\delta \rho$ that leads to $F[\rho + \delta \rho] = 0$ we must
solve,
\begin{equation}
J \cdot \delta \rho = - F[\rho]
\label{eqn:linsys.1}
\end{equation}
for $\delta \rho$. 

Computationally our tasks are to first evaluate the discrete form of
the Jacobian matrix, $\hat{J}$ and second to solve the discrete form of
Eq.(\ref{eqn:linsys.1}).  We numerically compute the Jacobian matrix by
perturbing the surface pointwise and examining the effect of of the
perturbation on the residual, $\hat{F}$. Let $\bar{\mu}$ denote
``independent'' points in the computational mesh, $\hat{\cal S}$. By
independent we mean the unique points (points modulo boundary identification)
on $\hat{\cal S}$.  In particular from the identifications made earlier,
there are $N_s^2 - 2N_s + 2$ independent points in $\hat{\cal S}$. $N_s
= N_{\theta} = N_{\phi}$ points at each of the poles are treated as one
point. $\bar{\mu} = 1$ represents the $\theta=0$ point for all the
$N_{\phi}$ points($0 \le \phi < 2\pi$) and $\bar{\mu} = N_s^2 - 2N_s +
2$ is the $\theta =\pi$ point.  Eq.(\ref{eqn:linsys.1}) then becomes a
linear system of equations where $\hat{J}$ is a $(N_s^2 - 2N_s + 2)
\times (N_s^2 - 2N_s + 2)$ matrix and $\hat{F}$ and $\delta \hat{\rho}$
are vectors of length $N_s^2 - 2N_s + 2$.  The $\bar{\mu}\bar{\nu}$
component of the Jacobian is then computed by perturbing $\rho$ at the
$\bar{\nu}$-th point and computing the change in the residual,
$\hat{F}$ at the $\bar{\mu}$-th point. Using a first order forward
difference approximation we have, 
\begin{equation}
\hat{J}_{\bar{\mu}\bar{\nu}} = \frac{1}{\epsilon} \left\{
\hat{F}_{\bar{\mu}}[\hat{\rho}_{\bar{\mu}} + \epsilon] -
\hat{F}_{\bar{\mu}}[\hat{\rho}_{\bar{\nu}}]       \right\},
\label{eqn:jacnumpert.1} \end{equation} where $\epsilon$ is the amount
by which we perturb the surface. We define $\epsilon$ to be the {\it
perturbation parameter}. The process for generating the components then
involves numerically evaluating $\hat{F}$ in only a small neighborhood
of the $\nu$-th point since $\hat{F}$ has a domain of dependence
dependent on the finite difference operators used.  In this case the
operators are (finite difference) derivative operators convolved with
interpolation operators.

The solution of Eq.(\ref{eqn:linsys.1}) is achieved via Newton's method.
We solve for the change $\delta \hat{\rho}$ that leads to a new surface 
$r = \hat{\rho}_{*}(\theta,\phi)$ that yields
$\hat{F}[\hat{\rho}_{*}] \sim 0$ up to $O(h^2)$, where $h$ is the Cartesian 
stencil spacing, proportional to $\delta \theta$. Newton's method then 
involves updating $\hat{\rho}(\theta,\phi)$ by $\delta \hat{\rho}$.
If $F$ were a linear operator then one iteration would result in a 
$\delta \hat{\rho}$ that leads to a solution. Since $F$ is nonlinear we have 
to iterate until the $L_2$-norm of the residual, $\|\hat{F}\|_2$, is driven 
down to a chosen stopping criterion. We discuss the implementation details
in the appendices and discuss further the properties of the Jacobian and 
Newton's method in the results section. We now turn our attention to the model
spacetime in which we shall conduct our numerical experiments.

\section{3+1 Splitting of the Kerr-Schild metric}

In the rest of the paper we focus on tests of the algorithm based on
boosted Schwarzschild and Kerr black holes. The particular form that we
use is given by the Kerr-Schild metric:  \begin{equation} g_{\mu \nu} =
\eta_{\mu \nu} + 2 H l_{\mu} l_{\nu}, \label{eqn:kerrschild.metric}
\end{equation} where $l_{\mu}$ is an ingoing null vector (i.e:  $g^{\mu
\nu} l_{\mu} l_{\nu} = \eta^{\mu \nu} l_{\mu} l_{\nu} = 0$), $H$ is a
scalar function of the spacetime  coordinates and $\eta_{\mu \nu}$ is
the Minkowski spacetime metric. This metric describes the Kerr and
Schwarzschild spacetimes. We note that under a Lorentz transformation
the spacetime metric is form invariant. By definition such a
transformation takes $\eta_{\mu \nu} \rightarrow \eta_{\mu \nu}$ and
$l^{\mu}$ and $H$ are transformed to a new null vector and left
unchanged (though evaluated at the new coordinate labels for the same
event) respectively.  This property makes our analysis easier since a
3+1 decomposition of Eq.(\ref{eqn:kerrschild.metric}) has the same form as 
a 3+1 decomposition of the boosted metric. As we shall
see, we only need to specify $H$, $l_{\mu}$ and their spacetime
derivatives in order to obtain the 3-metric and extrinsic curvature on
$\Sigma$.

For a vacuum spacetime, $l^{\mu}$ is geodesic and in the Kerr and
Schwarzschild spacetimes is the tangent to geodesics of ingoing
photons. The null nature of Eq.(\ref{eqn:kerrschild.metric}) leads to a
slicing of these spacetimes that is well behaved at the horizon. That
is, spacelike slices penetrate the horizon and hit the black hole
singularity. This is a desirable property for black hole excision in
computational applications and this metric has shown itself to be a good
choice for the study of single and multiple black hole evolutions with
exicision.

For the Kerr spacetime, $H$ and $l_{\mu}$ are given by
\begin{equation}
H = \frac{M r^3}{r^4 + a^2 z^2}
\label{eqn:kerr.hdefn}
\end{equation}
and 
\begin{eqnarray}
l_{\mu} = \left( 1, \frac{rx + ay}{r^2 + a^2}, 
\frac{ry - ax}{r^2 + a^2}, \frac{z}{r}\right),
\label{eqn:kerr.ldefn}
\end{eqnarray}
where $r$ is given by
\begin{equation}
\frac{x^2 + y^2}{r^2 + a^2} + \frac{z^2}{r^2} = 1,
\end{equation}
or
\begin{equation}
r^2 = \frac{1}{2}\left(\rho^2 - a^2 \right) + \sqrt{\frac{1}{4}
\left(\rho^2 - a^2 \right)^2 + a^2 z^2}.
\end{equation}
$M$ is the mass of the Kerr black hole and $a = J/M$ is the angular
momentum of the black hole and $\rho = \sqrt{x^2 + y^2 + z^2}$.

In the $a\rightarrow 0$ limit we get the Schwarzschild metric in ingoing
Eddington-Finkelstein coordinates where
\begin{equation}
H = \frac{M}{r},
\label{eqn:schw.hdefn}
\end{equation}
\begin{equation}
l_{\mu} = \left(1, \frac{x}{r}, \frac{y}{r}, \frac{z}{r} \right)
\label{eqn:schw.ldefn}
\end{equation}
and $r = \sqrt{x^2 + y^2 + z^2}$.

In a spacelike slice of either Kerr or Schwarzschild spacetimes the
apparent horizon is known to coincide with the intersection of the
event horizon with that slice. In the Kerr spacetime then the apparent
horizon is a surface of radius $r=r_{+}$:  
\begin{eqnarray} 
r_+ & = & M + \sqrt{M^2 - a^2} 
\label{eqn:kerr.horizon.location}
\end{eqnarray} 
and area 
\begin{eqnarray} 
A & = & 4\pi (r_+^2 + a^2).  
\end{eqnarray} 
In the more general nonstationary
case the apparent horizon and event horizon will not coincide. We use
the properties of the Kerr and Schwarzschild spacetimes to test out our
method for finding horizons.

To get the spacetime metric for a boosted black hole consider
$\bar{\cal O}$ to be the rest frame of the black hole, with coordinates
$(\bar{t},\bar{x}^i)$. Let ${\cal O}$ be another stationary frame with
coordinates $(t,x^i)$ such that ${\cal O}$ is related to $\bar{\cal O}$
via a Lorentz boost along the $\hat{\bf v} = (\hat{v_x}, \hat{v_y},
\hat{v_z})$ direction: in the ${\cal O}$ frame the black hole moves in
the $\hat{\bf v}$ direction with boost velocity, $v$ ($\delta_{ij}
\hat{v}^i \hat{v}^j = 1$).  As usual, we define $\gamma = 1 /
\sqrt{1-v^2}$.  
$H(x_{\bar{\mu}})$ and $l_{\bar{\mu}}$ (bar denoting $\bar{\cal O}$
frame) now transform as 
\begin{eqnarray} H(x_{\mu}) =
H(\Lambda^{\bar{\nu}}_{\mu} x_{\bar{\nu}}) 
\end{eqnarray} 
and
\begin{eqnarray} 
l_{\mu} = \Lambda^{\bar{\nu}}_{\mu} l_{\bar{\nu}}
(\Lambda^{\bar{\sigma}}_{\gamma} x_{\bar{\sigma}}) .
\end{eqnarray} 
These preserve the form of (\ref{eqn:kerrschild.metric}).

\subsection{3+1 Decomposition}

The standard ADM 3+1 form of the spacetime metric is given by
\begin{equation}
ds^2 = -\alpha^2 dt^2 + 
\gamma_{ij} \lb dx^i + \beta^i dt \rb \lb dx^j + \beta^j dt \rb
\label{eqn:3+1.form}
\end{equation}
If we compare (\ref{eqn:kerrschild.metric}) to (\ref{eqn:3+1.form}) and use 
the property that $l^{\mu} l_{\mu} = 0$, we find that the lapse is given by
\begin{eqnarray}
\alpha & = & \frac{1}{\sqrt{1 + 2 H l_t^2}},
\label{eqn:kerrschild.lapse}
\end{eqnarray}
and the shift is given by
\begin{equation}
\beta_i = 2 H l_t l_i
\label{eqn:kerrschild.shift}
\end{equation}
or
\begin{equation}
\beta^i = 2 H l_t \delta^{ij} l_j /(1+2 H l_t^2).
\label{eqn:kerrschild.shift}
\end{equation}
The 3-metric is given by
\begin{equation}
\gamma_{ij} = \eta_{ij} + 2Hl_i l_j.
\label{eqn:kerrschild.3metric}
\end{equation}
as expected and the extrinsic curvature is determined from
\begin{eqnarray}
K_{ij} &=& - \partial_t \gamma_{ij} / 2 \alpha + D_i \beta_j + D_j \beta_i \\
       &=& - \partial_t(H l_i l_j)/ \alpha + 2\left(D_i(H l_t l_j) + 
           D_j(H l_t l_i)\right)
\label{eqn:extcurv}
\end{eqnarray}
and 
\begin{eqnarray}
\gamma^{ij} & = & \delta^{ij} - 2 H \delta^{il} \delta^{jk} l_l l_k / 
(1 + 2 H l_t^2 ).
\end{eqnarray}
Note that 
\begin{eqnarray}
det \gamma_{ij} & = & 1 + 2 H l_t^2.
\end{eqnarray}

To obtain the 3-metric and extrinsic curvature we need to specify $H$,
$l_{\mu}$, $\partial_{\mu} H$ and $\partial_{\mu} l_{\nu}$ and substitute
these into Eq.(\ref{eqn:kerrschild.3metric}) and Eq.(\ref{eqn:extcurv}).
In order to 
evaluate $\partial_\mu l_\nu$ and $\partial_\mu H$ a specific choice of 
spacetime has to be made. 
For example for the Kerr spacetime we take the expressions for $H$ and
$l_{\mu}$ from Eq.(\ref{eqn:kerr.ldefn}) and Eq.(\ref{eqn:kerr.hdefn}),
compute their derivatives and substitute. This gives us a  ``Kerr-Schild''
slice of the Kerr spacetime.

\section{Results}

In the presentation that follows we first conduct a series of basic
tests  of the algorithm using metric and extrinsic curvature data 
just presented.  In the second part of this section we set up a 
3-dimensional Cartesian grid, $\hat{\Sigma}$, of $n^3$ points, on which we 
define a coordinate system where the black hole (either Kerr or 
Schwarzschild) 
is placed at the origin. Using 
Eq.(\ref{eqn:kerrschild.3metric},\ref{eqn:extcurv}) 
again we generate $\gamma_{ij}$ and $K_{ij}$ on $\hat{\Sigma}$ everywhere but 
the region that contains the curvature singularity (for Schwarzschild at 
$x^2+y^2+z^2=0$ and Kerr $\rho = \sqrt{x^2+y^2+z^2} \le a$).
  With the data on $\hat{\Sigma}$, we start the apparent horizon
locator with an initial guess surface which is a 2-sphere of radius
$r_0$. The horizon locator surface mesh sizes used in this calculation
are $N_s=33,41,49,65,81,97$. The stopping criterion for
the Newton iterations was determined empirically to be $10^{-9}$. The
cases we present  are (1) $v=0$, $a=0$ (unboosted Schwarzschild), (2)
$v=0$, $a \ne 0$ (unboosted Kerr), (3) $v\ne 0$, $a=0$ (boosted
Schwarzschild) and (4) $v \ne 0$, $a \ne 0$ (boosted Kerr).

\subsection{Tests with Eddington-Finkelstein metric data}
\label{section:eddfink.tests}

First, with $\hat{v}_x = 0$, $\hat{v}_y = 0$, $\hat{v}_z = 0$ and $a = 0$
(unboosted black hole in ingoing Eddington-Finkelstein coordinates) we show
some basic tests of the apparent horizon locator. Most importantly we show
that solutions obtained with our locator are $O(h^2)$. 
With these data all components of $\gamma_{ij}$ and $K_{ij}$ are non-zero. The 
latter property makes this a good initial model problem to work with, because
the computation is fully exercised in an analytically tractable situation.
As stated earlier the apparent horizon is expected to be located at $r=2M$.
These tests are conducted with data specified analytically where required.

\subsubsection{Residual Evaluation and Second Order Convergence} 

\begin{figure}[h] 
\begin{center} 
\leavevmode
\epsffile{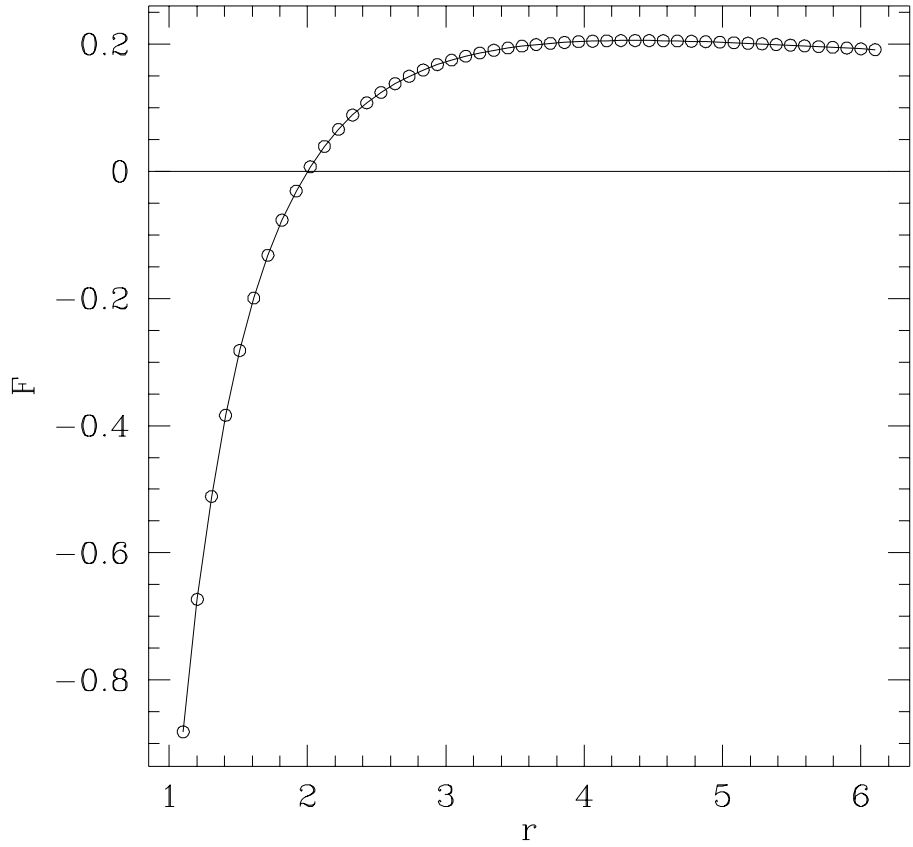} 
\end{center} 
\caption{Figure showing
the value of $F$ , given by eqn(\ref{eqn:residual.efa}), is shown as a
function of $r$. $F(r)$, is evaluated on a series of 2-spheres of radius $r_0$.
At $r=2M$ $F=0$ and as $r\rightarrow \infty$,
$F\rightarrow 0$. As $r\rightarrow 0$, $F\rightarrow -\infty$. The
maximum value of $F$ is $0.206/M $ at $r=4.372M$} 
\label{fig:analytic.f}
\end{figure}

\begin{figure}[h]
\begin{center}
\leavevmode
{\epsffile{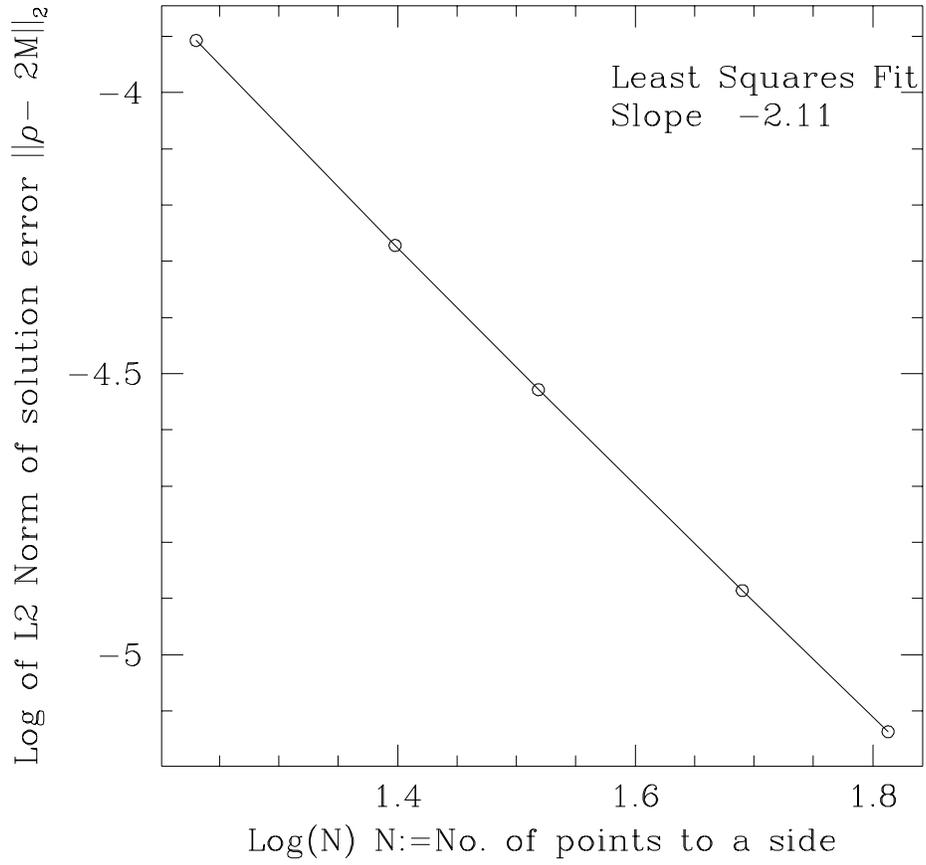}}
\end{center}
\caption{This graph shows the log of the error norm of the numerical solution,
$\|\hat{\rho} - \rho\|_2$ versus $log(N_s)$. The slope of the graph is
$-2.11$ and hence the solution is $O(h^2)$.}
\label{fig:ef.soln.conv}
\end{figure}

We place the black hole at the origin of the computational
domain($x=0,y=0,z=0$). In spherical symmetry for this metric the apparent 
horizon equation becomes the algebraic equation, 
\begin{equation} 
F(r) = \frac{1-2M/r}{r\sqrt{1+2M/r}} =0.
\label{eqn:residual.efa} 
\end{equation} 
A plot of $F(r)$  is shown in FIG.(\ref{fig:analytic.f}).  
At $r=2M$ we have $F=0$. A useful test of the evaluation of the expansion of 
the outgoing
normals $F(r)$ is to see if indeed the residual  $\hat{F}[\hat{\rho}]$
is correctly evaluated to $O(h^2)$ as
\begin{equation} 
\hat{F} = F + e_2 h^2 + \ldots , 
\end{equation} 
where $e_2 h^2$ is the leading
order truncation error term.  Given that the exact value is known
for $\hat{F}[\hat{\rho}]$ we can approximate the leading order truncation
error.  We carry out a convergence test by
evaluating $\hat{F}[\hat{\rho}]$ on a 2-sphere of $r = 2M$ for a series
of mesh sizes, $N_s=17,25,33,49,65,96,129$.  
We examined the behavior of $\log{\|\hat{F}\|_2}$ (where this is the $L_2$ 
norm) versus $\log{N_s}$, where $N_s$ is the number of mesh points on one side 
of the $N_s\times N_s$ mesh.  At $r=2M$ $\hat{F} \sim e_2 h^2$ and so the 
$L_2$-norm, $\|\hat{F}\|_2 \sim \|e_2\|_2 h^2$. Since $h \propto 1/N_s$ we 
expect that if the residual is $O(h^2)$ then the slope of a plot of
$\log{\|\hat{F}\|_2}$ versus $\log{N_s}$ should be 
$-2.0$, which we validated via a least squares fit. 
A closely related test is to also evaluate $\log{\|\hat{\rho} - \rho\|_2}$ 
versus $\log{N_s}$, where $\hat{\rho}$ is the numerical solution from the 
apparent horizon locator and $\rho$ is the exact horizon location.  
FIG.(\ref{fig:ef.soln.conv}) shows the result. From a least squares 
fit to a straight line the slope is found to be $-2.1$ which validates our
solution as $O(h^2)$.

\subsubsection{Jacobian}

\begin{figure}[h]
\begin{center}
\leavevmode
{\epsffile{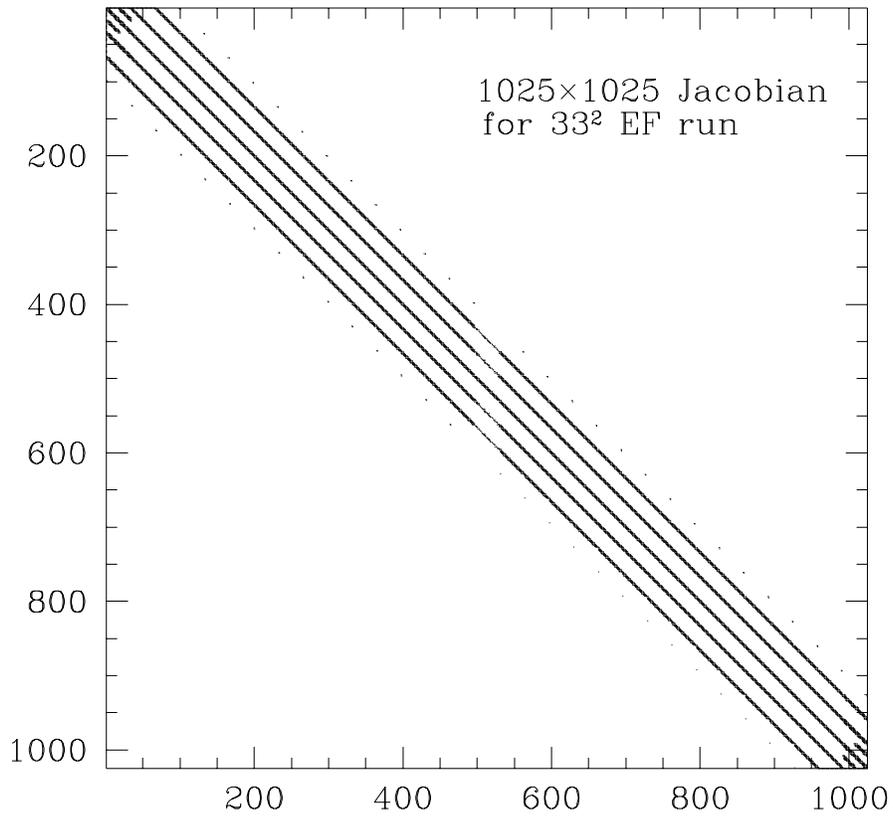}}
\end{center}
\caption{This figure shows the sparsity structure of the Jacobian matrix
for a $33\times 33$ run.}
\label{fig:sparsity}
\end{figure}

For the same 2D mesh discussed we generate the Jacobian matrix for a single
Newton step.
FIG.(\ref{fig:sparsity}) shows the structure of the matrix for a 
$33\times 33$ run. 
There are $1025$ {\it independent points} on $\hat{S}$ and hence 
$J_{\bar{\mu}\bar{\nu}}$ is a $1025\times 1025$ matrix. The dots in the figure 
are non-zero Jacobian entries. There are seven bands in this matrix with 2 
additional ones in the vicinity of the poles at $\bar{\mu}=1$ and 
$\bar{\mu}=1025$. The structure reflects the domain of dependence of 
the finite difference operators used in the evaluation of $\hat{F}$. Here it
comes from a combination of interpolations and 
Cartesian finite differencing. Near $\bar{\mu}=1$ and $\bar{\mu}=1025$ the 
additional bands come from our special choice of interpolation stencils at the
poles, as discussed in the appendix.

The structure reflects the fact that a perturbation at a single mesh
point affects the residual in a small neighborhood around it so we can
optimize the generation of the Jacobian to $O(N)$ by evaluating
$\hat{F}[\rho + \delta \rho]$ only in a small neighborhood of the perturbed
point.
The Jacobian generation was found to be order $O(N^p)$ where $p =1.08$
and $N$ is the total number of independent
points on the 2D computational mesh.

A matrix $A$ is defined to be diagonally dominant \cite{Ortega} if its 
elements, $A_{ij}$, satisfy
\begin{equation}
\sum^{n}_{\stackrel{j=1}{j \neq i}} \mid A_{ij} \mid \le \mid A_{ii} \mid 
\mbox{for all i}.
\label{eqn:dominance}
\end{equation}
We found that the Jacobian is {\it not} diagonally
dominant since the inequality in Eq.(\ref{eqn:dominance}) is not satisfied 
for all $i$ and $j$. 

This is of interest  since for some iterative solution techniques 
(Gauss-Seidel and SOR for example) a sufficient condition for the solution of 
a linear system, $A\cdot x = b$, is that the matrix, $A$, be diagonally
dominant. In our case we concluded from early experiments that indeed
such simple iterative solvers did not converge for this problem.

The Jacobian matrix is not symmetric but it is well-conditioned for the 
spacetimes that we have 
considered. For a $33\times 33$ run the Jacobian has a condition number of
$\kappa$ of about $10^4$ to $10^5$ where,
\begin{equation}
\kappa = \|A\| \|A^{-1}\|,
\label{eqn:cond.number}
\end{equation}
(In their definition of the condition number, Dongarra 
et al.\cite{dongarra} use the $L_1$-norm.)  The condition number tells us how
close the matrix $A$ is to being singular. A very large condition number or 
a reciprocal condition number close to machine epsilon tells us that
$A$ is singular. An identity matrix has a condition number of $1$. To
estimate $\kappa$ we used the {\tt LINPACK } library routine, {\tt DGECO}.

\subsubsection{Solution of the Linear system}

As stated before to locate the apparent horizon using our technique we have
to obtain a solution, $\delta \hat{\rho}$, to the linear system
\begin{eqnarray}
\hat{J}\cdot \delta \hat{\rho} = -\hat{F}[\hat{\rho}] \nonumber
\end{eqnarray}
which is the discrete form of Eq.(\ref{eqn:linsys.1}).

Since the properties of the matrix do not allow us to use the standard 
iterative methods such as Jacobi and Gauss-Seidel methods, 
We use a modified conjugate-gradient method 
due to Kershaw\cite{kershaw} (The standard form of the Conjugate gradient 
method will not work since $J$ is not symmetric.) Kershaw's method, termed the 
Incomplete LU-Conjugate gradient method (ILUCG), can solve any linear system,
$A\cdot x = b$, with $A$ being any nonsingular, sparse matrix. The method
involves preconditioning the matrix via an incomplete LU decomposition.
This method has worked quite well for our purposes. In principle other schemes
for solving the resulting system can be used. 

\subsubsection{Solution for the apparent horizon location} 
Using the ILUCG method to solve for $\delta \hat{\rho}$ we demonstrate the
Newton solver's ability to sucessfully locate apparent horizons
in the Eddington-Finkelstein metric data. The apparent horizon in this data
is a 2-sphere of radius $2M$.
Using a 2-sphere of radius
$r_0$, centered on the origin as the initial surface we carried out a
series of runs for $r_0=0.5..3.0$ with a $33 \times 33$ mesh. 
Table~(\ref{table:ef.soln}) shows the radius, $r_0$, of the initial starting 
2-sphere, the number of Newton iterations taken if it converged, the final 
residual value and the solution error. 
\begin{table}
\caption{This table shows a series of runs carried out for various initial
conditions. The initial surface is a 2-sphere of radius $r_0$ where $r_0$
is shown in the first column. The number of iterations taken for the solution
to achieve $\|\delta \rho\|_2 < 10^{-10}$ is given. Note that the final
error in the solution remains a constant provided the solver is able to drive
$\|\delta \rho\|_2$ below the specified stopping criterion. The perturbation
parameter used to generate the Jacobian was $10^{-5}$.}
\begin{center}
\begin{tabular}{ccccc}
\hline
$r_0$ & \# iterations & Final $\|\hat{F}\|_2$ & $\|\hat{\rho}-\rho\|_2$ & \% error \\
\hline
0.5  & 10    &    2.2E-10 & 3.0E-05 & 1.5E-03 \\
0.75 & 9     &    2.5E-10 & 3.0E-05 & 1.5E-03 \\
1.00 & 9     &    9.7E-11 & 3.0E-05 & 1.5E-03 \\
1.25 & 8     &    9.5E-11 & 3.0E-05 & 1.5E-03 \\
1.50 & 7     &    1.6E-10 & 3.0E-05 & 1.5E-03 \\
1.75 & 7     &    9.6E-11 & 3.0E-05 & 1.5E-03 \\
2.00 & 4     &    9.7E-11 & 3.0E-05 & 1.5E-03 \\
2.25 & 7     &    9.8E-11 & 3.0E-05 & 1.5E-03 \\
2.50 & 8     &    1.0E-10 & 3.0E-05 & 1.5E-03 \\
2.75 & 9     &    9.5E-11 & 3.0E-05 & 1.5E-03 \\
3.00 & 12    &    9.9E-11 & 3.0E-05 & 1.5E-03 \\
\hline
\end{tabular}
\end{center}
\label{table:ef.soln}
\end{table}
The stopping criterion is that the norm of the change in the solution, 
$\|\delta \hat{\rho}\|_2$ be less than $10^{-10}$.
We see that for all the cases, provided the solver managed to drive 
$\|\delta \rho\|_2$ below $10^{-10}$ the final percentage error remains
fixed; what differed in each case was the number of iterations taken
and rate of convergence. 
Once the solver
has driven $\hat{\rho}$ into the vicinity of the solution the Newton 
convergence is quadratic. This happened in the 6th iteration in the $r_0=0.5$
run. For the $r_0=2.0$ run the solver took four iterations to converge down
to the stopping criterion. 
In another series of runs with $N_s=33$, $M=1$ and $r_0=2.5$ the perturbation
parameter, $\epsilon$, was varied from $10^{-1}$ to $10^{-6}$.
An optimum value of $\epsilon$ for this
metric data was found to be about $10^{-4}$ to $10^{-6}$. In general the 
stopping criterion need not be as stringent as we have set it. In numerical 
spacetimes where the metric data will have truncation error associated with 
them the truncation error of $F$ is expected to be much larger than our
test stopping criterion.
In that case a larger stopping criterion should to be chosen to avoid wasting
computational effort. 
The optimum value is dependent on the error in the metric data.
In the results we present in the paper however we drive the residuals down 
as far as possible.  On the other hand, a perturbation parameter $\epsilon$ 
must be chosen such that 
$\hat{F}[\hat{\rho} + \epsilon] - \hat{F}[\hat{\rho}]$ is sufficiently large
that this expression is not dominated by truncation error.

\subsubsection{Numerical Metric data}

Tests described so far used data analytically computed at each point as needed. 
Since the ultimate goal is to incorporate this apparent horizon location
algorithm into an evolution code it is useful to gauge the performance of
the algorithm with numerical metric data and with the data 
structures expected in the real application, where, for 
instance, part of the domain is excised from consideration.  Thus we 
set up the same Eddington-Finkelstein data on a 3D Cartesian grid of 
$n^3$ points, with a region of this grid excised to emulate the situation in 
an evolution code where the interior of the black hole is excluded.
The apparent horizon surface which is embedded in this 3D Cartesian grid
typically does not lie on Cartesian grid points and as a result an 
interpolation tool is required. If the surface mesh, during the course of the 
Newton iterations, overlaps the excised region then extrapolation is required. 
We make use of an interpolator/extrapolator written by 
S. Klasky\cite{klasky_interpolators}
to obtain the 3-metric, extrinsic 
curvature and the spatial derivatives of the 3-metric at any point. The use 
of an interpolator brings in truncation errors associated with the 
interpolation/extrapolation operations. In the following we show that even 
with extrapolation errors the solver works quite well in locating
apparent horizon surfaces. 

We set up an uniform 3D Cartesian grid, $\hat{\Sigma}$, of size $n^3$. On this 
grid we excise a region interior to a sphere of radius $R_m$ centered at 
$(x_m,y_m,z_m)$ so that the metric data is defined for $r > R_m$ and 
undefined for $r < R_m$ where r is the Cartesian distance 
in Kerr-Schild coordinates from the excision center.

In the following discussion on radial and offset apparent horizon locations 
we take $n=65$ for the Cartesian grid 
(with $h=1/8$) and $N_s=33$ for the surface mesh.  The stopping 
criterion used in the horizon finder is $\beta = 10^{-4}$. That is, if 
$\|\delta \rho\|_2 < \beta$ 
then the Newton iterations are stopped.
The perturbation parameter, $\epsilon$, is taken to be $10^{-4}$. The 
interpolator tool is used to fourth order. The initial guess surface used 
is a sphere of radius $r_0 = 2.1M$ centered at the origin of the Cartesian 
grid. With these parameters we carry out two set of tests. 
The first is a radial test of the horizon locator with the use of the 
interpolator and the second is an offset test. These tests examine the effect 
of extrapolation of metric data on the residual, $\hat{F}$, and solution to 
the apparent horizon equation.

\subsubsection{Apparent horizon location (Radial tests with excision)}

\begin{figure}[h]
\begin{center}
\leavevmode
{\epsffile{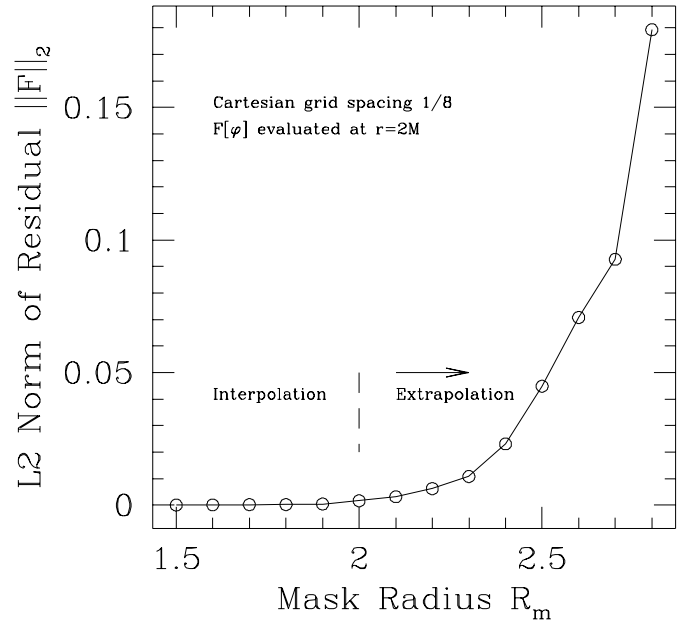}}
\end{center}
\caption{Graph of the $L_2$-norms of the residual,$\|\hat{F}[\varphi]\|$,
versus the mask radius, $R_m$}
\label{graph:evalnorms.1}
\end{figure}

\begin{figure}[h]
\begin{center}
\leavevmode
{\epsffile{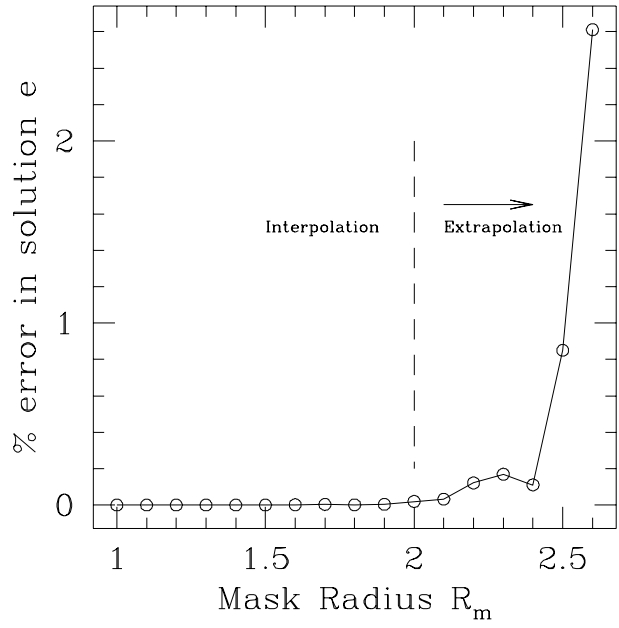}}
\end{center}
\caption{Graph of percentage relative error in $\rho$  versus mask radius.
Note that
past $R_m=2.5$ the solver did not meet the stopping criterion of $10^{-4}$.}
\label{graph:locate.1}
\end{figure}

In the first test case we center the black hole at $ \lb 0,0,0 \rb $.
The masked region is also centered at $ \lb 0,0,0 \rb$. We carry out a
series of tests with the excision radius, $R_m$, varying from $1.5$ to
$2.6$.  Thus the apparent horizon is in the defined region ($R_m < 2M$)
for some of the tests, and for others it is inside the excised region
($R_m > 2M$). This provides evidence of the effect of extrapolations on
the residual of the apparent horizon equation, $\hat{F}$, and the error
in its solution.  FIG.(\ref{graph:evalnorms.1}) illustrates the
behaviour of the $L_2$-norm of the residual,
$\|\hat{F}\|_2$, as a function of $R_m$. FIG.(\ref{graph:locate.1})
shows the 
percentage relative error of the solution of the apparent horizon equation as
a function of $R_m$. The percentage relative error is calculated from the
exact solution for $r=\rho(\theta,\phi)=\bar{\rho}=2M$. 
With this exact solution, we calculate the percentage relative error, 
$e\equiv \|\rho - \bar{\rho} \| / \bar{\rho} \times 100\%$.
For $R_m < 2M$ the derivative interpolator/extrapolator
uses interpolation for regions near the apparent horizon location ($r=2M$),
while for $R_m > 2M$ it uses extrapolation. As $R_m$ increases further 
the errors due to extrapolation increase, as expected. 
This can be seen in FIG.(\ref{graph:evalnorms.1}) 
where $\|\hat{F}\|_2$ increases quickly for
$R_m \agt 2.4$, as does the error shown in  FIG.(\ref{graph:locate.1}).
At $R_m = 2.5M$, the solver could not bring  $\|\delta \rho\|_2$ 
down to below $10^{-4}$, and so failed to meet the stopping criterion.
This can be understood in terms of the Cauchy-Schwarz inequality
\cite{Ortega}, 
Since $\|{ J \cdot \delta \rho }\| = \|{ \hat{F}}\|$ we have that
\begin{equation}
\|{ \delta \rho}\| \ge \frac{\|{ \hat{F}}\|}{\|{J}\|}.
\label{eqn:cond}
\end{equation}
At $R_m = 2.2$ where $\| \hat{F}\| \sim 10^{-3}$ and
$\|\delta \rho\| \sim 10^{-5}$, we have from Eq.(\ref{eqn:cond}) 
that $\|J\| \sim 10^2$. Therefore at
$R_m = 2.5M$ we expect with $\|\hat{F}\|\sim 10^{-2}$  
that $\|\delta \rho\| \sim 10^{-4}$. By relaxing the criterion 
past $R_m =2.5$ we can still obtain a solution.
Past $R_m=2.6$ the convergence progressively worsens. For example, at 
$R_m=2.9$, $\|\hat{F}\|$ could not be brought below $10^{-3}$, and 
the solution error is 5\%. 
The amount of error sustained from interpolation of the metric data is
dependent on the resolution of the Cartesian grid and the behaviour of
the functions being interpolated. If the gradients of $\gamma_{ij}$ and
$K_{ij}$ are large near the horizon then a larger interpolation error is
sustained. This in turn leads to a larger truncation error in $F$. In the
numerical evolution of black hole spacetimes with excision then buffer zones
may not be necessary for the location of apparent horizons. However, for
other reasons buffer zones might be necessary.

\subsubsection{Locating Offset apparent horizons}

We examine behaviour of the locator with the derivative interpolator for
a black hole offset so that it overlaps the excised region. This is
important in tracking moving black holes\cite{mhuq.texas.talk}.

\begin{figure}[h]
\begin{center}
{\epsffile{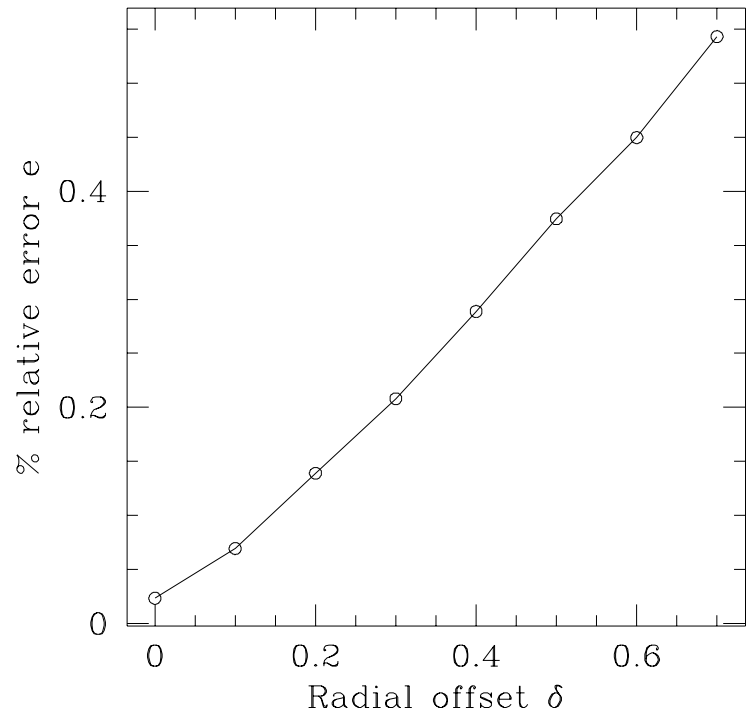}}
\end{center}
\caption{Percentage relative error as a function of the offset $\delta$.}
\label{fig:offset}
\end{figure}

The center of the masked region is at $ \lb 0,0,0 \rb$ 
and the black hole of radius 2M is centered at ($\delta/\sqrt{3}, 
\delta/\sqrt{3},\delta/\sqrt{3}$), so that the radial
distance between the mask center and the hole is $\delta$. With a grid 
spacing of $h=1/8$, an offset of $\delta=1$ corresponds to approximately 8 
grid zones.
FIG.(\ref{fig:offset}) shows the percentage relative error in the apparent
horizon location as a function of the offset $\delta$. As the graph 
illustrates, up to $\delta=0.7$ the percentage relative error is below one 
percent. (At $\delta =0.7$ the percentage error is 0.6\%.) From 
$\delta=0.7$ onwards, however the solver becomes sensitive to initial 
conditions and extrapolation errors and quickly ceases to converge.

At $\delta=0.7$, about 5-6 grid points offset, we are still able to find
horizons. 
Generally in
explicit time-evolution codes the CFL 
condition \cite{CFL.paper} restricts the black hole motion from one time slice
to another, to be less than one zone ($\delta < h$ or about $\delta \sim 0.1$
in our test case). Hence we expect, based on the results for our model 
spacetime as shown in FIG.(\ref{fig:offset}), that in such an evolutionary 
scheme with a similar resolution we will be able to track black hole 
apparent horizons to less than 0.1\%.

\subsection{Apparent Horizons in Boosted Kerr data}

\begin{figure}[h]
\begin{center}
{\epsffile{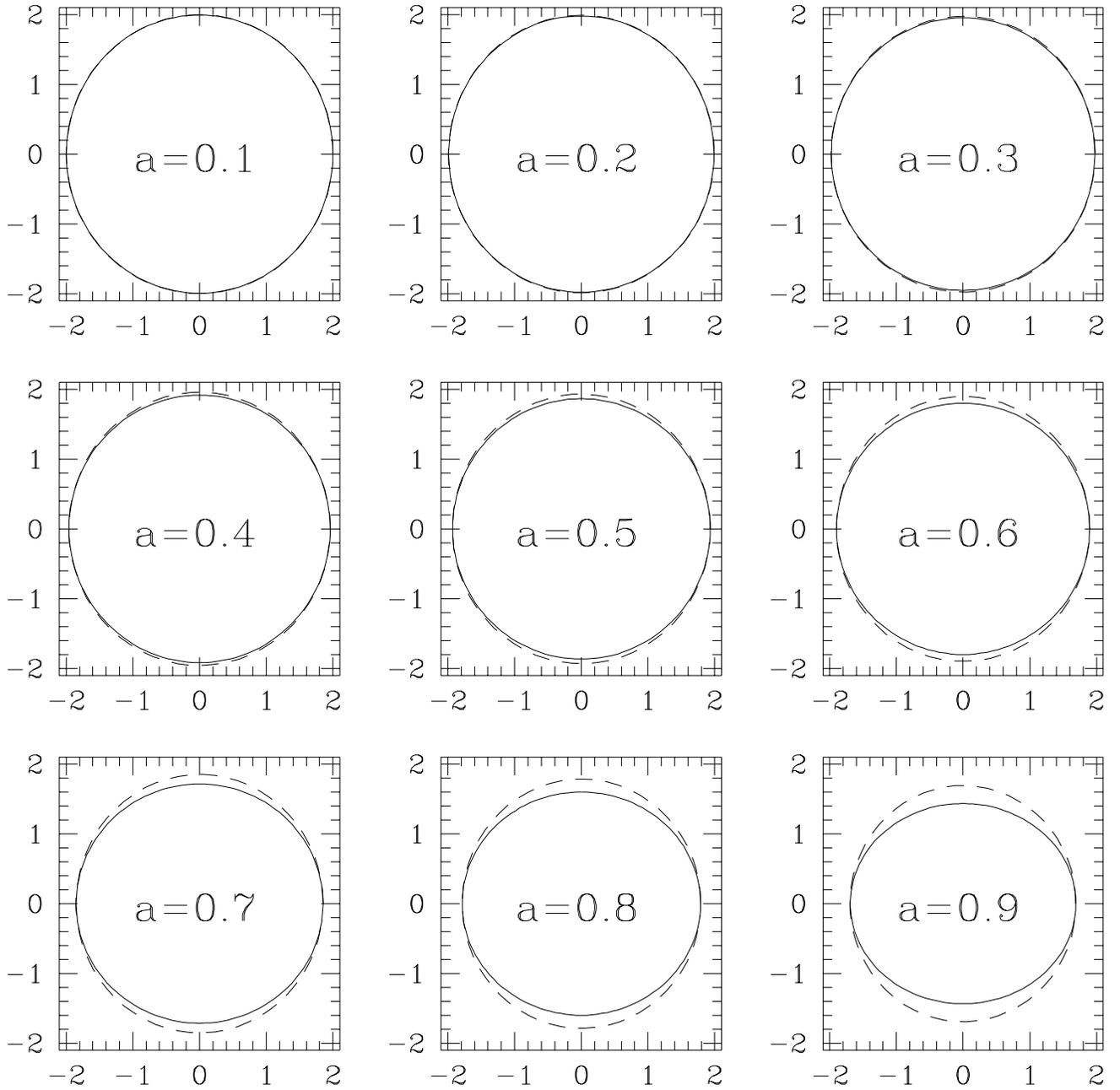}}
\end{center}
\caption{The following figure shows the cross-section of apparent horizons
located for various values of $a$. The solid line shows the $\phi = \pi$
slice of the apparent horizon. The dashed line shows the $\theta=\pi/2$ slice
of the horizon. As expected the $\phi=\pi$ slices show increasing deformation
for increasing $a$.}
\label{fig:kerr.range}
\end{figure}

In this section we now focus on apparent horizon location for boosted
Kerr and Schwarzschild black holes.  For the data that follow we excise a 
2-sphere of radius $r > a$ centered about the origin from the computational 
domain to avoid the ring singularity structure of the Kerr black hole. Using 
the interpolator tool in conjunction with the apparent horizon locator, we 
locate horizons for various values of the angular momentum parameter, $a$. 
The horizon locator begins with a trial surface which 
is a 2-sphere of radius $r_0=2M$. The locator was run for
$a=0.0, 0.1, 0.2, \ldots , 0.9$ at $N_s=33$. FIG.(\ref{fig:kerr.range}) 
shows a cross-section of the horizon in the $xz$ plane as a function of $a$. 
The apparent horizon is seen to have the shape of an oblate spheroid.
In the runs used to generate these data we used $\epsilon = 10^{-5}$ and
a stopping criterion that ensured that the $l_2$-norm of $\hat{F}$ on
the computational mesh was less than $10^{-11}$.

\begin{figure}[h]
\begin{center}
\leavevmode
{\epsffile{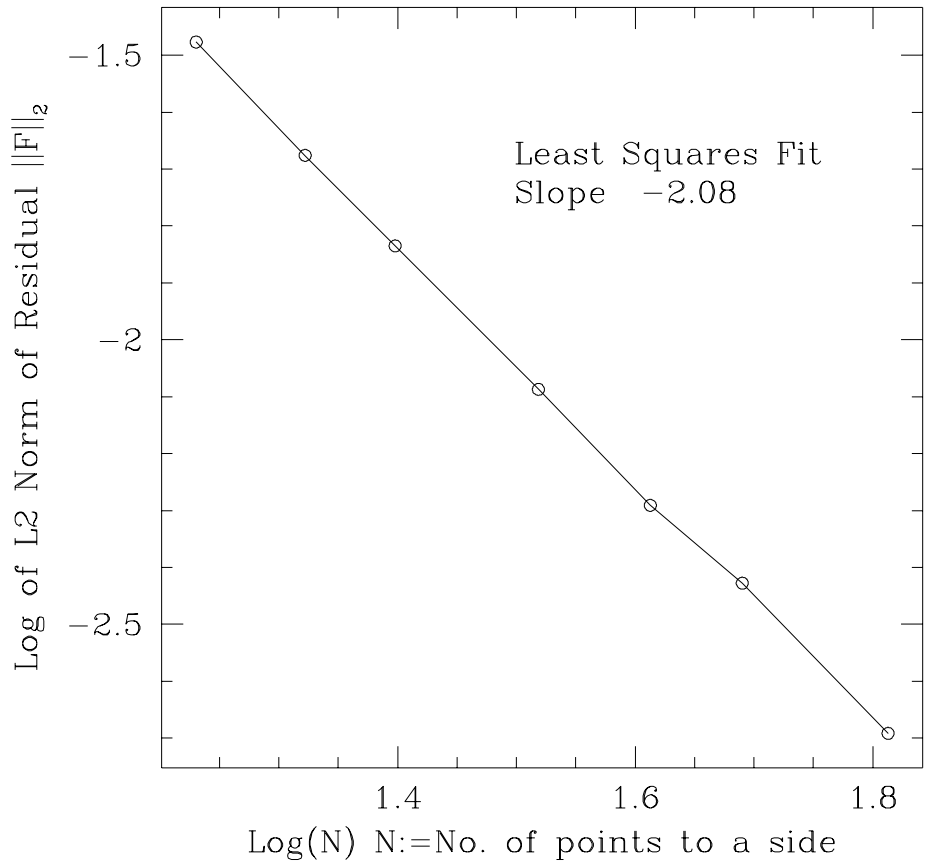}}
\end{center}
\caption{$\|\hat{r}-r_+ \|_2$ versus $N_s$ for $a=0.9$.}
\label{fig:kerr.convergence}
\end{figure}

FIG.(\ref{fig:kerr.convergence}) shows the $l_2$-norm of the error in the 
solution, $\|(\hat{r} - r_+)/r_+ \|_2$, versus mesh size. This set of runs 
was carried out with $a=0.9$ and $N_s = 17,25,33,49,65$.
Where $r_+$ is given by (\ref{eqn:kerr.horizon.location}) and the 
$\hat{r}$ is 
computed from
\begin{equation}
\hat{r} \equiv \left\{ \frac{1}{2}\left(\hat{\rho}^2 - a^2 \right) +
\left[\frac{1}{4}\left(\hat{\rho}^2 - 
a^2\right)^2 + a^2 z^2\right]^{1/2}\right\}^{1/2}
\end{equation}
where $\hat{\rho}=\sqrt{x^2 + y^2 + z^2}$ is the solution from the apparent 
horizon locator. 
From a least squares fit the slope is found to be $-2.1$ and the solution
is $O(h^2)$.

The area of the event horizon in the Kerr spacetime\cite{MTW} is given by
\begin{eqnarray}
\cal{A} 
        & = & 4\pi\left( r_+^2 + a^2 \right)
\label{eqn:kerr_area}
\end{eqnarray}

\begin{figure}[h]
\begin{center}
{\epsffile{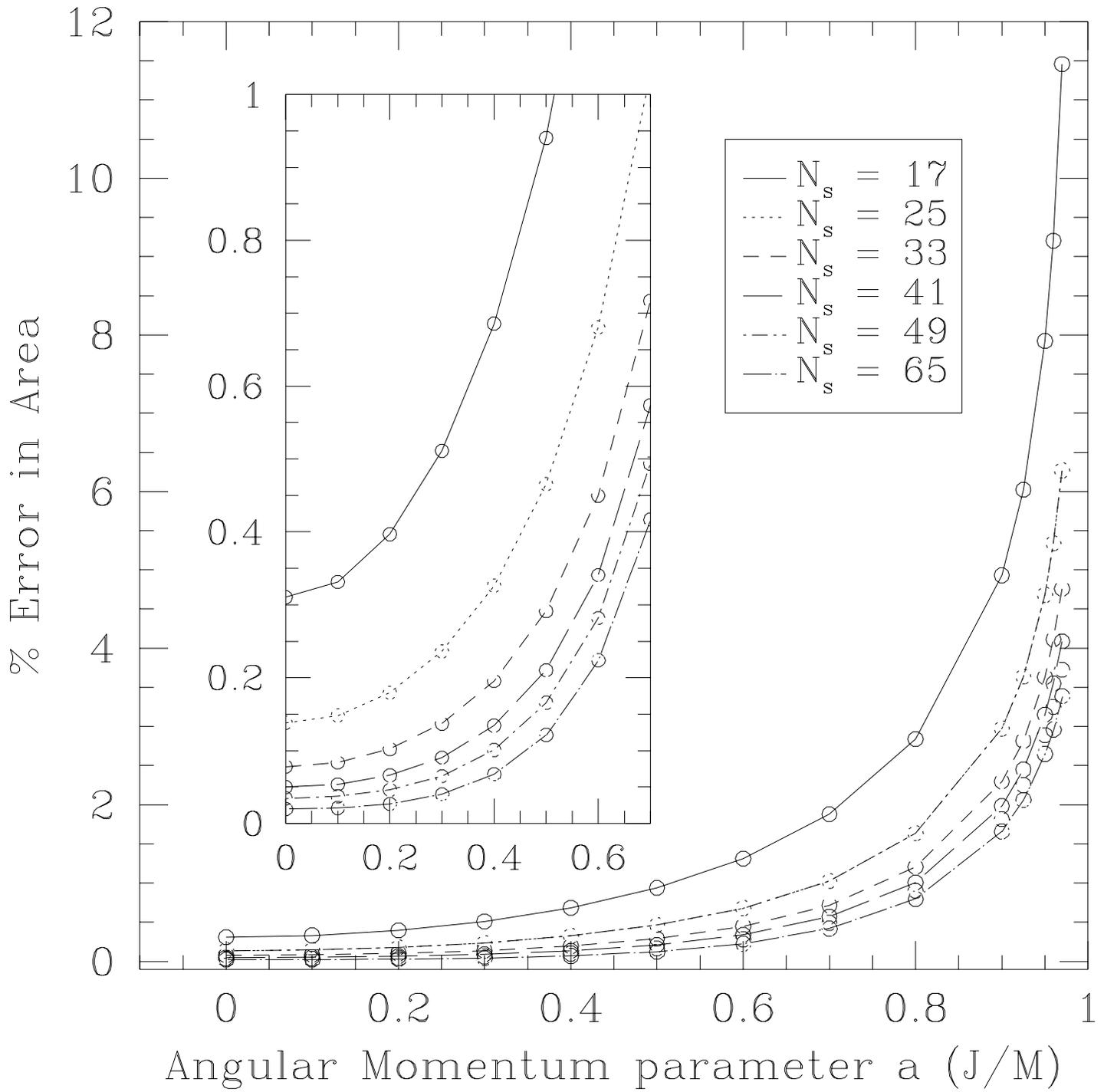}}
\end{center}
\caption{The percentage error in the area of the apparent horizon
for a Kerr hole versus $a$ is shown for $N_s=17,25,33,41,49,65$. }
\label{fig:kerrareatest}
\end{figure}

\begin{figure}[h]
\begin{center}
\leavevmode
{\epsffile{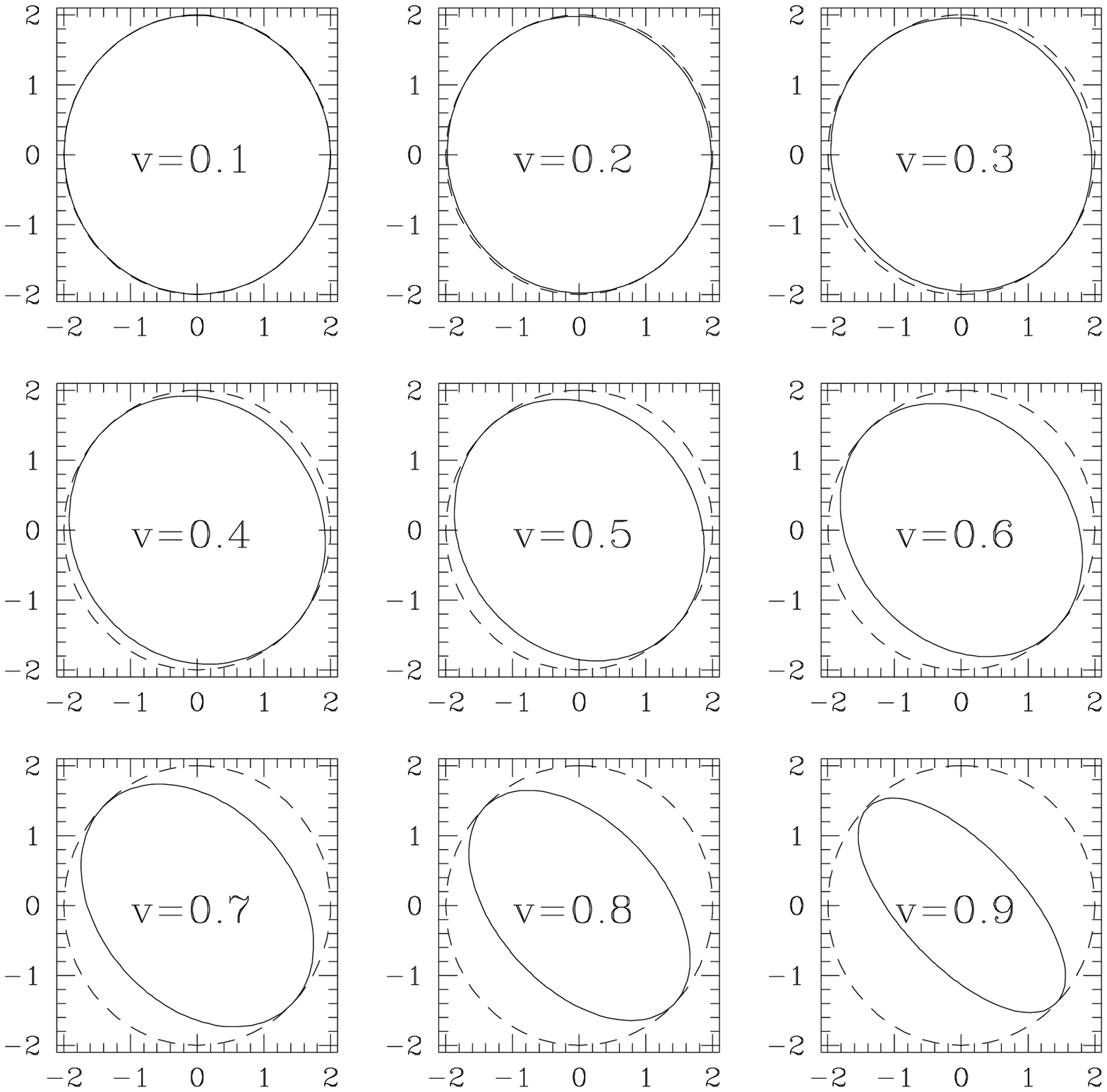}}
\end{center}
\caption{This figure shows the y-z cross-section of Schwarzschild black hole
apparent horizons
located for boost velocities of $v=0.0,0.1,...,0.9$ in the y-z direction.
The dashed circle in each of these figures is the apparent horizon for an
unboosted Schwarzschild black hole. Note that the point of contact is along
the y-z direction orthogonal to the direction of the boost.}
\label{fig:schwbst9}
\end{figure}

\begin{figure}[h]
\begin{center}
{\epsffile{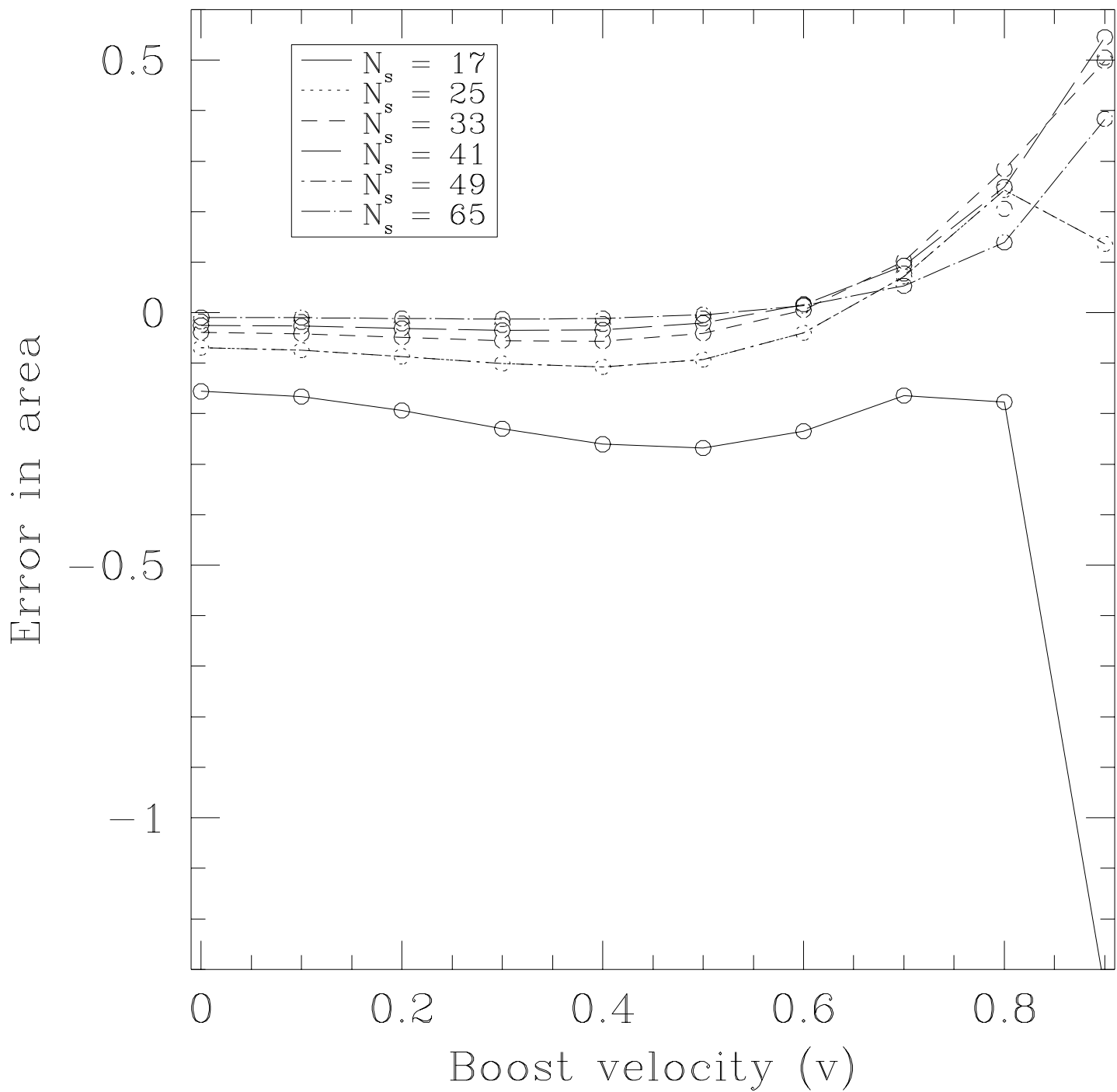}}
\end{center}
\caption{This figure shows the error in the areas of apparent horizons found
for a Schwarzschild black hole boosted in the y-z direction.}
\label{fig:schw_yz_errareas}
\end{figure}

\begin{figure}[h]
\begin{center}
{\epsffile{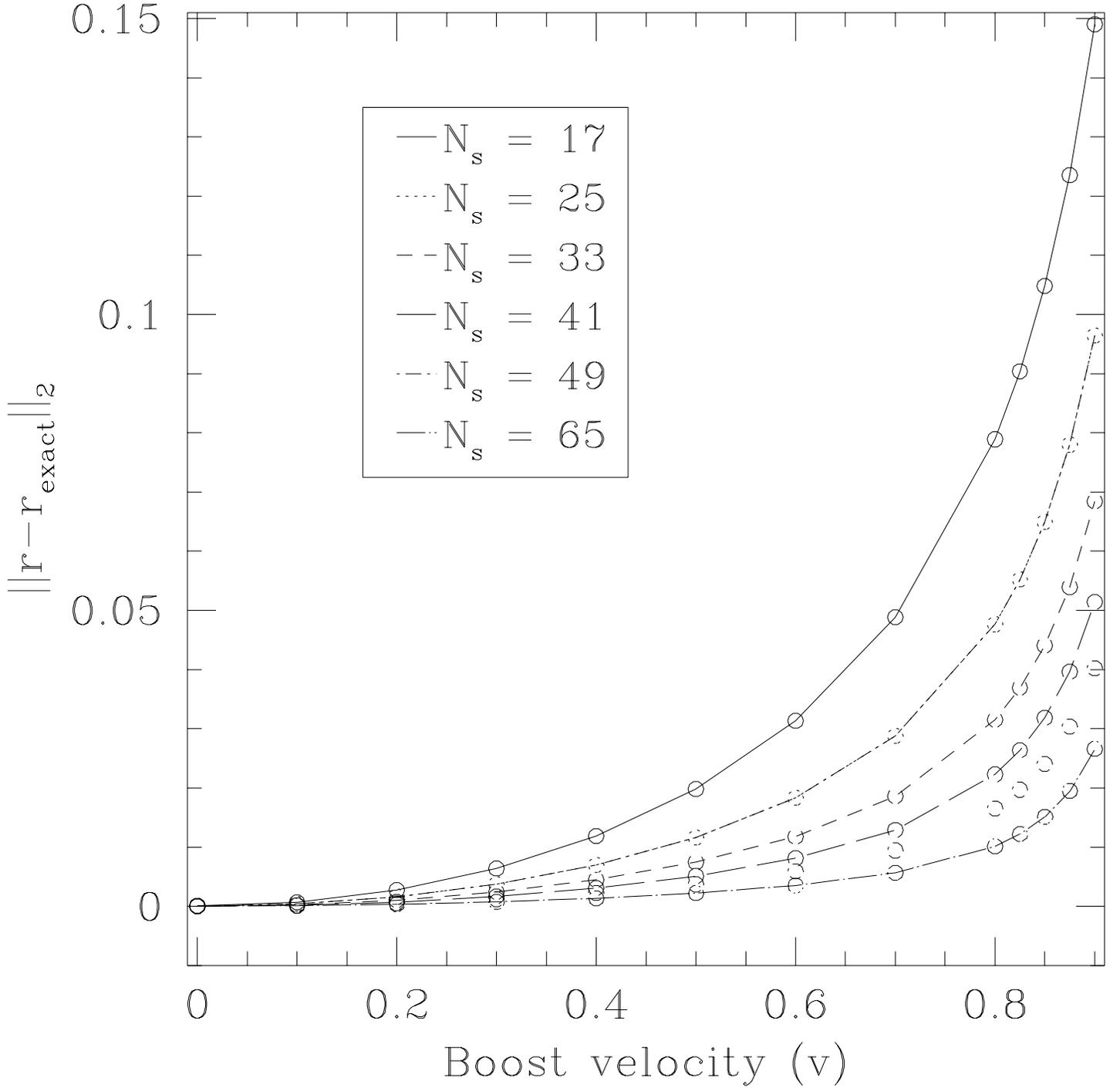}}
\end{center}
\caption{This figure shows the error in the radial coordinate
location $r=2M$ for a Schwarzschild black hole apparent horizon
for $v=0.0,..,0.9$. The black hole is boosted in the xyz-direction.}
\label{fig:schwbst.rrerror}
\end{figure}

Let $\hat{\cal A}$ be the computed apparent horizon area.
FIG.(\ref{fig:kerrareatest}) shows the percentage errors
$\|(\hat{\cal A} - {\cal A})/{\cal A} \times 100\%\|_2$ versus $a$ for 
various resolutions $N_s=17,\ldots, 65$. 
The area of the apparent horizon is computed via a technique which projects
the 3-metric, $\gamma_{ij}$, onto the 2-surface to obtain an area element
$\sqrt{{}^{(2)}\gamma} d\theta d\phi$, and then computes the surface integral.
FIG.(\ref{fig:kerrareatest}) shows the percentage errors in the area for
increasing resolution. 
We now consider Schwarzschild and Kerr black holes boosted in the yz-direction.
That is, we look at $\hat{v}_x=0,\hat{v}_y=1/\sqrt{2}, \hat{v}_z=1/\sqrt{2}$ 
and $a=0,0.9$.
In each of these cases we locate apparent horizons for $v=0,0.1,\ldots, 0.9$.
From $v=0$ to $v=0.8$ we started with a two-sphere of radius $2M$ and found
an apparent horizon with outgoing expansions driven down to $10^{-12}$. For
$v>0.8$ we had difficulty driving the expansions down. As a result we 
utilized the solution at $v=0.8$ as an initial guess and were subsequently
able to find horizons by stepping every $0.25$ from $v=0.8$ to $v=0.9$.
We used $\epsilon = 10^{-5}$ again for these runs. At $v=0$ the initial guess
is the apparent horizon and there within the six Newton iterations the 
expansions were driven down around $10^{-12}$. The first Newton iteration
took the expansions down around $10^{-6}$. For $v=0.5$ starting from an
initial guess of a sphere of radius $\rho=2M$ it took four Newton iterations to 
drive the expansions down around $10^{-6}$ and nine Newton iterations to get
down to $10^{-12}$. Typically in a numerical time-evolution of such a 
spacetime we would not need to drive the expansions down to this level. 
If we are utilizing
a surface within the apparent horizon as an excision boundary then we need
only to drive the expansions down far enough to be certain an apparent horizon
is present. FIG.(\ref{fig:schwbst9}) shows the yz-cross-section of the
apparent horizon for various boost velocities compared against an unboosted
black hole apparent horizon cross-section. We find that the apparent horizon
is Lorentz contracted in the yz-direction in the boosted coordinates. 
We have considered a slice of such a boosted spacetime in 
which the event horizon appears Lorentz contracted in the resulting 
coordinates. We know that in these spacetimes the apparent horizon should
coincide with the event horizon and we find that this is indeed the case.
First, the area of the apparent horizon coincides with the area of the 
event horizon which is invariant under a boost. 
FIG.(\ref{fig:schw_yz_errareas}) shows the error in the apparent horizon
area as a function of $v$ for various resolutions. We find that with increasing
resolution the error in the area converges towards zero. This demonstrates that
the area of the apparent horizon is invariant under a Lorentz boost.
This is coupled to an interesting property of the Kerr-Schild type of 
metrics that $r=r_+$ remains fixed for Kerr and Schwarzschild black holes.
This is illustrated in FIG.(\ref{fig:schwbst.rrerror}) where we show
the error in the radial coordinate $r=2M$ on the apparent horizon
for various boost velocities. In this case the black hole is boosted in 
the $xyz$-direction for generality. That is, 
$\hat{v}_x=1/\sqrt{3},\hat{v}_y=1/\sqrt{3},\hat{v}_z=1/\sqrt{3}$ and $a=0$.
Here $r$ is computed from the boosted coordinates. 
We find that $r$ converges towards $2M$ for 
increasing resolution satisfying yet another property of the boosted 
Kerr-Schild spacetime.

FIG.(\ref{fig:kerrbst0_surf}) shows surface plots of the apparent horizon
for $v=0,0.3,0.6$ and $v=0.9$ displayed in Kerr-Schild Cartesian coordinates. 
Note how distorted the apparent horizon
gets with increasing boost velocities. As seen in the figures for the 
$yz$-boosts the boosted apparent horizons in this case are always contained
within the apparent horizon for $v=0$. That is, the boost contracts the
apparent horizon in the boost-direction.
Again for a boost velocity of $v=0.5$ it took the solver eight Newton
steps to drive the expansions down around $10^{-12}$. The stopping criterion
used in this run was $10^{-12}$ and the final expansions are $\sim 10^{-13}$.
On average it took four Newton steps to drive the expansions down to
$\sim 10^{-6}$  and nine Newton steps to $\sim 10^{-12}$ starting from an 
expansion of $0.1$.

\begin{figure}[h]
\begin{center}
{\epsffile{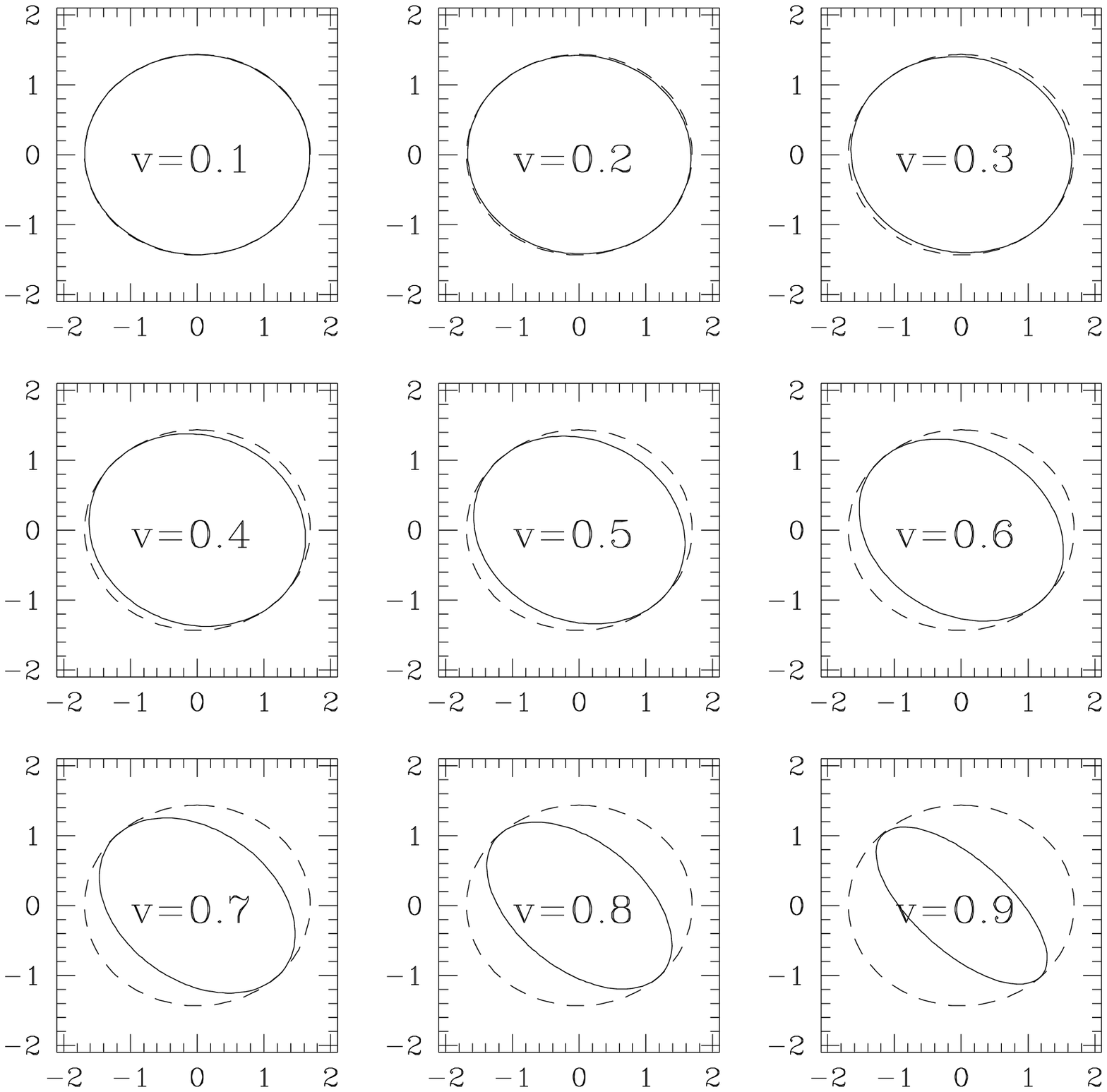}}
\end{center}
\caption{This figure shows the y-z cross-section of  Kerr (a=0.9) black hole
apparent horizons
located for boost velocities of $v=0.0,0.1,...,0.9$ in the y-z direction.
The dashed circle in each of these figures is the apparent horizon for an
unboosted Schwarzschild black hole. Note that the point of contact is along
the y-z direction orthogonal to the direction of the boost.}
\label{fig:kerr_yz_boost_slice}
\end{figure}

\begin{figure}[h]
\begin{center}
{\epsffile{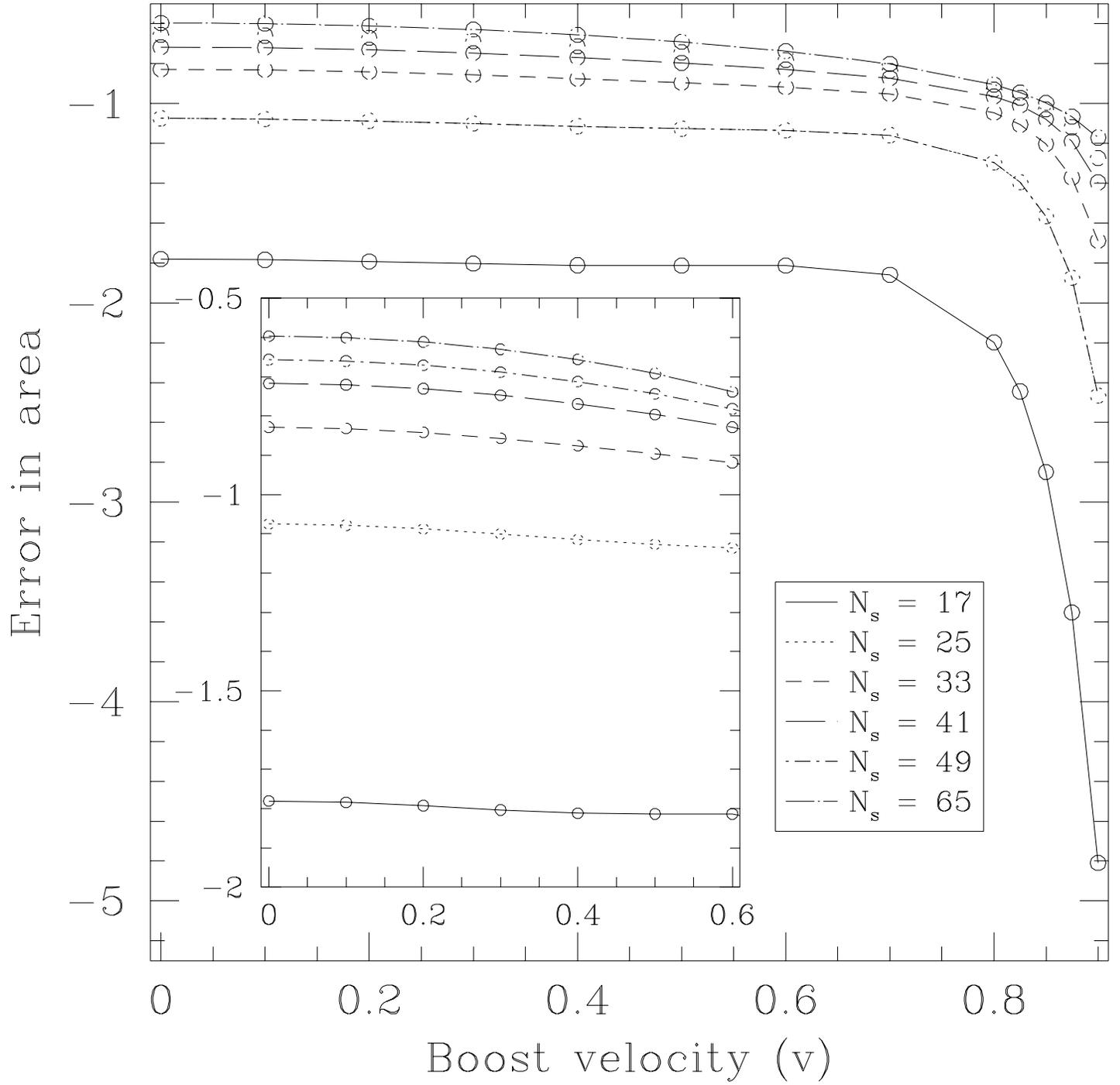}}
\end{center}
\caption{This figure shows the error in the areas of apparent horizons found
for a Kerr black hole boosted in the y-z direction with $a=0.9$.}
\label{fig:kerr_yz_errareas}
\end{figure}

In the case of a boosted Kerr black hole with $a=0.9$ the results are again
very similar to those of the Schwarzschild black hole. Note that now with
$a=0.9$ and $v\rightarrow 0.9$ we get even more distorted apparent horizons.
These results show that this algorithm for finding apparent horizons does
quite well with such large distortions. In addition the cost of finding 
these surfaces increases by only two additional Newton steps.
FIG.(\ref{fig:kerr_yz_boost_slice})
shows the $yz$-cross-sections for the apparent horizon found for $a=0.9$
as a function of $v$. Again the boosted apparent horizon is contained within
the unboosted one and Lorentz contracted. 
FIG.(\ref{fig:kerr_yz_errareas}) shows the error in the area for the same
data. With $a=0.9$ we expect that the area should be $\sim 36$. The graph
shows that for increasing resolution the error in the area tends towards zero.
Hence the area remains fixed with increased boost velocity as is expected.

\begin{figure}[h]
\begin{center}
{\epsffile{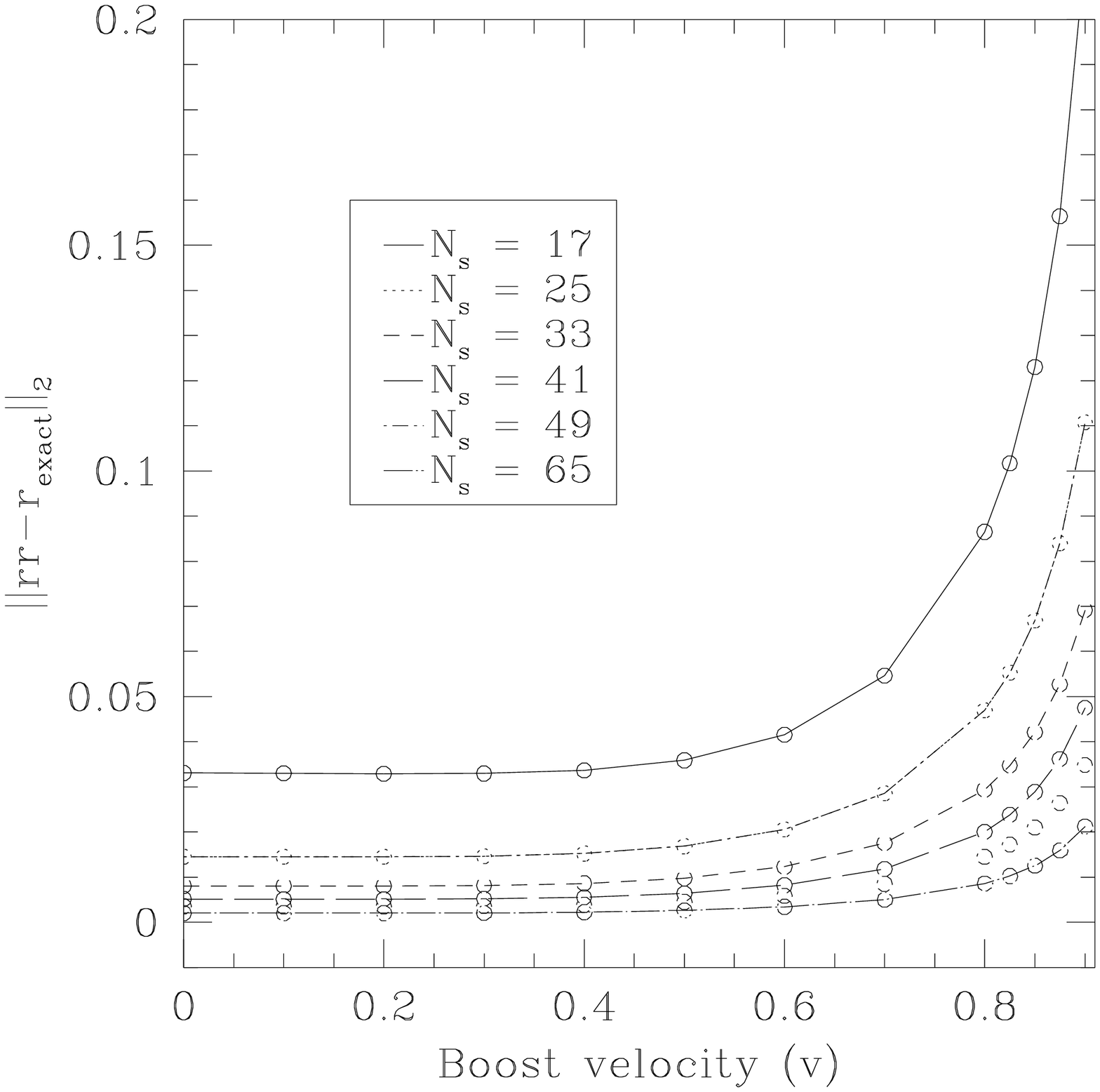}}
\end{center}
\caption{This figure shows the error in the spheriodal radial coordinate
location $r=r_+$ for a Kerr black hole apparent horizon ($a=0.9$)
for $v=0.0,..,0.9$. The black hole is boosted in the xyz-direction.}
\label{fig:kerrbst.rrerror}
\end{figure}

Similarly, $r$ computed from the boosted coordinates remains fixed at $r_+$
as is shown by FIG.(\ref{fig:kerrbst.rrerror}). The apparent horizons found
here were obtained with $\hat{v}_x=1/\sqrt{3}, \hat{v}_y=1/\sqrt{3}, 
\hat{v}_z=1/\sqrt{3}$ and
$a=0.9$. That is, the boost was in the $xyz$-direction with magnitude $v$.
Again we find that $r$ on the apparent horizon converges to $r_+ \sim 1.4$
with increasing resolution for all boost velocities. At a resolution of
$33 \times 33$ we have an percentage error of $8\%$ and $1\%$ at 
$65 \times 65$. 
FIG.(\ref{fig:kerrbst9_surf}) shows surface plots of the apparent 
horizon for the boosted Kerr black hole for $v=0,0.3,0.6$ and $0.9$.
Note how distorted the final apparent horizon surface is. Our algorithm
required one more Newton step to drive the expansions
down to $10^{-13}$ for $v=0$ compared to $v=0.9$. It took six Newton 
iterations
to drive the resolution from about $0.2$ at the inital step to $10^{-6}$ for
both boost velocities. Hence, this algorithm has the advantage that given
sufficient resolution on the computational mesh, the work done does not
drastically increase for increasing distortions.

\section{Discussion}

We have demonstrated in this paper that our method based on finite difference
techniques is viable for locating very distorted boosted Kerr black hole 
apparent horizons.
We have shown that the located horizons obey the expected analytical
rule, of invariance of the area of the event horizon, in cases
corresponding to at-rest or boosted single black holes, where the
apparent horizon is known to coincide with the event horizon. We have
additionally given a number of computational tests demonstrating the
behavior of the tracker on interpolated or extrapolated data which is
realistically similar to that from evolutions. In other contexts algorithm has 
been thoroughly tested with the canonical set of test problems such as the two 
and three black hole initial data sets and additionally in an
evolution code tracking the apparent horizon for a Schwarzschild black hole
in geodesic slicing\cite{mhuq.phd} and demonstrated to be capable of tracking
apparent horizons in boosted Schwarzschild data\cite{mhuq.texas.talk}. 
Those tests and the tests given here show its viability as a method for 
locating 
black hole apparent horizons and using them for black hole excision. Since
black hole excision is essential for long-term evolutions of 
single or multiple black hole spacetimes.  It is very useful to have 
efficient apparent horizon locators that can locate apparent horizons 
``fast'' relative to the time taken for an evolution time step. 

Our algorithm is dominated primarily by computations of the Jacobian matrix
in the use of Newton's method. These operations are optimized such that they
scale as ${\cal O}(N)$ approximately where $N$ is the total number of points
on the two-dimensional mesh used for the solution. With a $33\times 33$ mesh
we find that each Newton iteration takes on average $20$ Origin 2000 CPU 
seconds. This is independent of the distortion of the apparent horizon. 
However the number of Newton iteration steps is determined by
the ``distance'' of the initial starting surface from the final solution. During
the course of an evolution it is expected that the apparent horizons over
several timeslices will be ``close'' enough to each other that two to three
Newton iterations will be sufficient to locate the horizon at low accuracy 
with the
expansion of the outgoing null rays on its surface being at the level of
$10^{-5}$ or $10^{-6}$. Obtaining a better accuracy requires more Newton
iterations and the number of timesteps taken depends also on the
accuracy of the background metric data. Typically in our model problems
eight iterations will take us below $10^{-10}$. 

One of the drawbacks of a Newton's method for finding apparent horizons is
its sensitivity to the initial guess. An initial guess outside of the radius
of convergence will not lead to a solution. Additionally Newton methods are
known to be sensitive to high frequency components in the solution. This is
demonstrated in axisymmetry by Thornburg\cite{thornburg.1}. Sensitivity to
the initial guess can be easily handled by combining the Newton method 
algorithm with
apparent horizon trackers that are based on flow methods. The flow finder is
used to obtain an initial guess for the Newton method which then converges
on the solution very quickly. 

The efficiency of our boundary-value method can be compared to the
efficiency of other approaches (variations of flow mothods) due to Tod,
as developed by Shoemaker et al.\cite{shoemaker} and fast flow methods 
developed by Gundlach\cite{Gundlach}.  The flow method is based on a
parabolic partial differential equation whose rate of convergence to the 
solution slows as it approaches it. Typically for a $33\times 33$ run the 
flow method takes on the order of thousands of seconds to converge down
to expansions of $10^{-4}$. The advantage of this method however is its
ability to find multiple apparent horizons from an arbitrary initial guess.
Combined with the Newton finder this will result in a robust apparent horizon
finding scheme.

We can also compare the effort to spectral decomposition methods.  We
concentrate on a method similar to that of Nakamura et al.\cite{nakamura.1} 
in which the equation for the apparent horizon surface is written:
\begin{eqnarray}
\rho(\theta,\phi) & = & \sum_{l=0}^{l_{max}} \sum^{l}_{m=-l} a_{lm}
Y_{lm}(\theta,\phi). 
\label{eqn:seriesexp_sph}
\end{eqnarray}

We do not have access to an apparent horizon finder based on pseudo-spectral
methods but we will analytically compute the coefficients for the case of 
a boosted Schwarschild black hole; this will give some insight into the range 
of harmonics required, and some idea of what scaling of these methods might
be.

In Kerr Schild coordinates, the hole, with boost in the
z-direction, has the shape of a spheroid,
\begin{eqnarray}
\frac{x^2}{a^2} + \frac{z^2}{b^2} = 1,
\label{eqn:ellipsoid}
\end{eqnarray}
where we have supressed the y-direction.

Notice that the axes $a$,$b$ of the ellipsoid obey $b^2/a^2 = 1-v^2$, which
demonstrates that the eccentricity is directly proportional to the boost
velocity, $\epsilon = v$, for this case. Hence even
ellipses with moderate ratio of axes, such as that for v=0.9, where the
ratio is a little less than 0.5, have moderately large eccentricities.
We will approximate the form Eq.(\ref{eqn:ellipsoid})  with an axisymmetric 
series of the form Eq.(\ref{eqn:seriesexp_sph})
(the general case would have nonaxisymmetric terms also).  We find 
it more convenient to work with Legendre polynomials than with 
the spherical harmonics directly.

Since we work with Legendre polynomials, we drop the y- coordinate in
the spheroid expression, to obtain :
\begin{eqnarray}
R(\theta) & = & b / \sqrt{1 - \epsilon^2 \sin^2 \theta }  \\
          & = & b / \sqrt{1 - \epsilon^2 ( 1- q^2) },
\label{eqn:spheroid_axi}
\end{eqnarray}
where $q = \cos \theta$.

To obtain the expansion of expression Eq.(\ref{eqn:spheroid_axi}) in terms of 
$P_m$, we first expand using the binomial theorem. 
\begin{eqnarray}
R(\theta)/b & = & \sum^{\infty}_{s=0} \left( 
\begin{array}{c} -1/2 \\ s \end{array} \right)
(-1)^s \epsilon^{2s} (1-q^2)^s.
\label{series2}
\end{eqnarray}
This converges for all $v < 1$.

Using the binomial theorem again for $(1-q^2)^s$ we substitute
\begin{eqnarray}
(1-q^2)^s & = & \sum^{s}_{r=0} 
\left( \begin{array}{c} s \\ r \end{array} \right) (-1)^r q^{2r}
\end{eqnarray}
in (\ref{series2}) to obtain
\begin{eqnarray}
R(\theta)/b & = & \sum^{\infty}_{s=0} 
\left( \begin{array}{c} -1/2 \\ s \end{array} \right)
(-1)^s \epsilon^{2s} 
\sum^{s}_{r=0}
\left( \begin{array}{c} s \\ r \end{array} \right)
(-1)^r a_{sr}
\label{series3}
\end{eqnarray}
where 
\begin{eqnarray}
a_{sr}(q) & = & 
\sum^{r}_{n=0} \frac{2^{2n} (4n+1) (2r)! (r+n)!}{(2r+2n+1)! (r-n)!} P_{2n}(q)
\end{eqnarray}
and we made the substitution
\begin{eqnarray}
q^{2r} & = & \sum^{r}_{n=0} 
\frac{2^{2n} (4n+1) (2r)! (r+n)!}{(2r+2n+1)! (r-n)!} P_{2n}(q).
\nonumber
\end{eqnarray}
By exchanging the summations over $r$ and $n$ and then $n$ and $s$ it is 
possible to rewrite Eq.(\ref{series3}) as
\begin{eqnarray}
r(\theta) = \sum_{n=0}^{\infty} C_{2n} P_{2n}(\theta),
\end{eqnarray}
where
\begin{eqnarray}
C_{2n} & = & 2^{2n} (4n+1) \sum^{\infty}_{s=n} 
\left( \begin{array}{c} -1/2 \\ s \end{array} \right)
(-1)^s \epsilon^{2s}  \times \nonumber \\
& & \sum^{s}_{r=n} 
\left( \begin{array}{c} s \\ r \end{array} \right)
\frac{(-1)^r(2r)! (r+n)!}{(2r+2n+1)! (r-n)!}.
\label{thecoeffs}
\end{eqnarray}

\begin{figure}[h]
\begin{center}
\epsffile{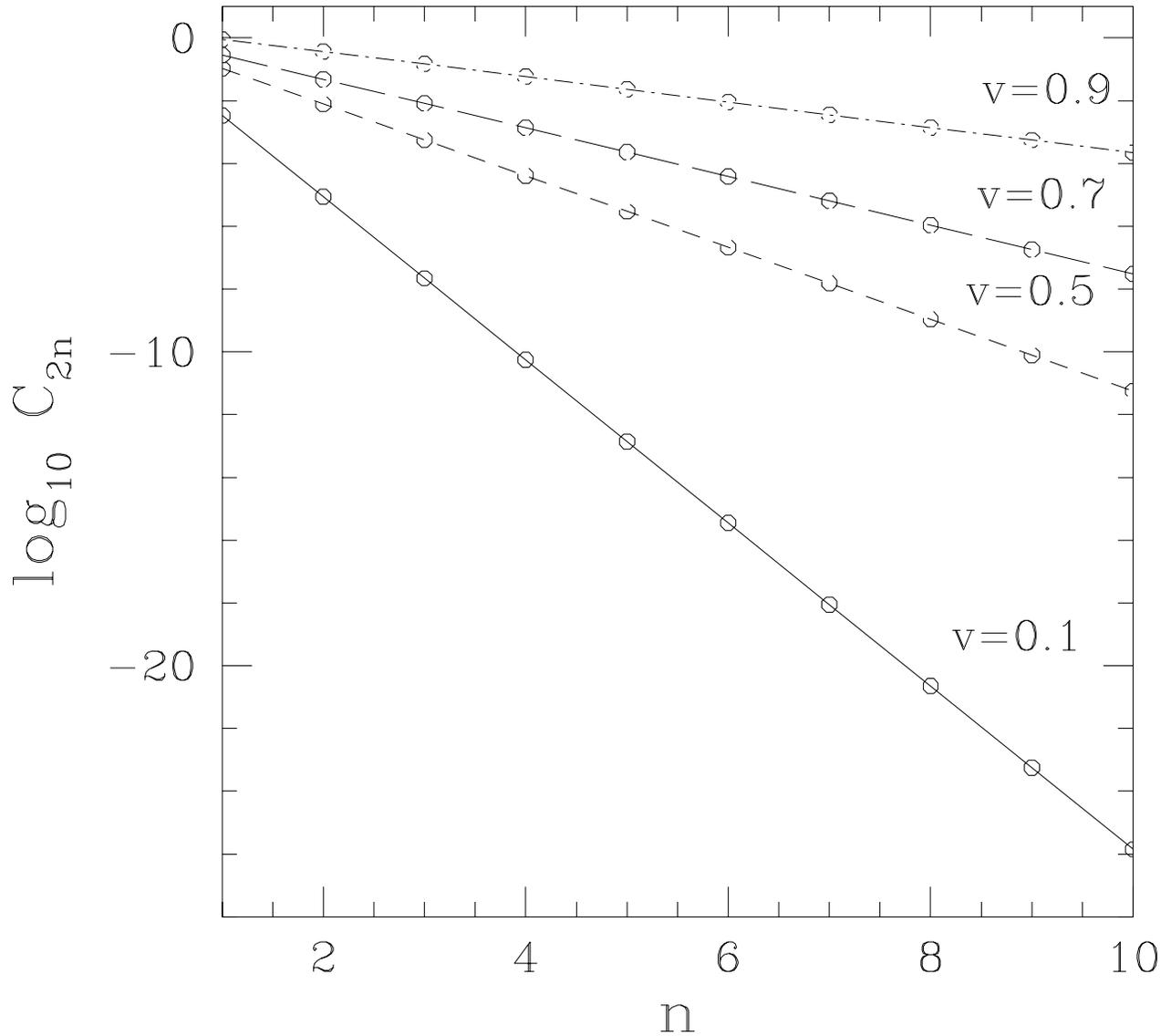}
\end{center}
\caption{$log_{10}(C_{2n})$ for $n=1$ to $10$ are shown for $v=0.1$,
$v=0.5$, $v=0.7$ and $v=0.9$. The coefficients scale as $e^{-k n}$ where
$k$ is determined by the shape of the horizon. For $v=0.1$ we have $k\sim 6$
and for $v=0.9$ $k\sim 0.9$.}
\label{fig:coeffs}
\end{figure}

FIG.(\ref{fig:coeffs}) gives the coefficients $C_{2n}$ for 
$n= 1,..., 10$, and for
several values of v.  While FIG.(\ref{fig:coeffs}) shows the exponential
convergence of algorithm with $n$, it also shows that the coefficient of the
convergence is small for $v \sim 0.9$.
It can be seen that the number of required terms
approaches 20 for $v=0.9$ if the error is required to be less than $10^{-3}$.  
(The general sum would have polynomials of odd as well as even order,
and for each l, a set of azimuthal quantum numbers spanning -2n to
2n).  Hence in general, to compute the distorted apparent horizon would
take a search over $20^2$ parameters in a minimization routine. This is
equivalent to inverting a full matrix of this size, and would be
expected to be slow. The boundary value problem is expected to be much
faster. It is a fact that the
boundary value problem as now implemented does not handle multiple
black-hole spaces, so its speed is counteracted by the impossibility of
using it in 2-hole cases.  However, it may be possible to use a 
flow method which does recognize the existence of seperate black holes,
run down to find the two holes each with some accuracy, and then to use
this boundary soluion code to quickly get a highly accurate result.  We
are confident such a combined tool would be of great utility.

\section{Acknowledgements}

We wish to thank S.Klasky for providing and developing the 3d interpolation 
tools used in a part of this work as well as for discussions and work on
applications of interpolation with excision. 
MFH wishes to thank J.Thornburg for stimulating discussions 
on apparent horizon location during the early part of this work.
This work was supported by NSF PHY9800722 and
PHY9800725 and PHY9800970.

\newpage
\section{Appendix A: Evaluation of the residual}
\label{section:apphor.eval}

\begin{figure}[h]
\begin{center}
  \epsffile{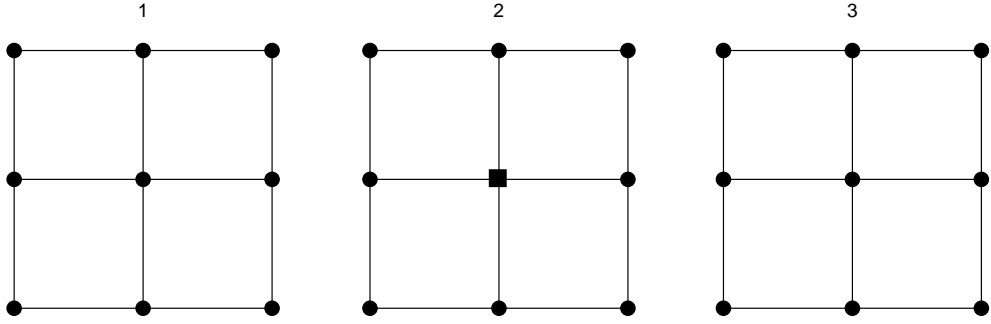}
\end{center}
\caption{This figure shows 3 planes of the finite difference molecule 
${\cal N}$. Plane 2 is the center plane with $z=z_0$. The center point is
denoted
by a filled square which is the mesh point $(x_0,y_0,z_0)$. Each of the other
stencil points in plane 2 are a distance $\pm\delta x$ and $\pm\delta y$
away and are denoted by circles. Plane 1 is at $z=z_0 -\delta z$ and
plane 3 is at  $z=z_0 + \delta z$.}
\label{fig:molecule.pp1}
\end{figure}

The evaluation of Cartesian derivatives on $\hat{\cal S}$ is carried out by
constructing 3D finite difference stencils at each mesh point on
$\hat{\cal S}$. The finite difference stencil, denoted by ${\cal N}$,
consists of 26 additional points around each mesh point. These 26 points are
$\pm\delta x$, $\pm\delta y$ and $\pm\delta z$ away from the central mesh
point as shown in FIG.(\ref{fig:molecule.pp1}). These points, as shown,
are organized into three planes of constant $z$: 
$z= z_0-\delta z, z_0, z_0 + \delta z$. Each plane
contains the nine nearest neighbors to the center point, including the center point itself in the case of $z=z_0$. We use a single
discretization scale $h$ ($\delta x = \delta y = \delta z = h$) which is
always proportional to the mesh spacing $\delta \theta = \pi /(N_s-1)$.

\begin{figure}[h]
\begin{center}
  \epsffile{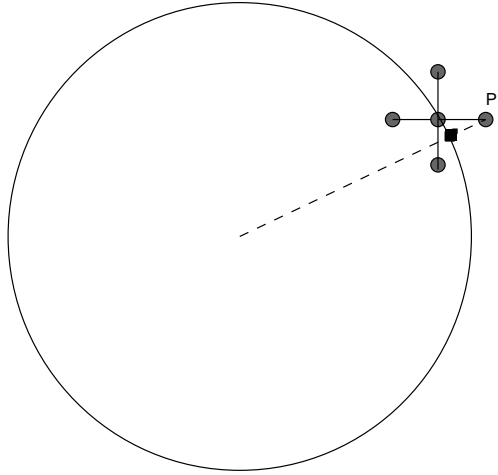}
\end{center}
\caption{P is the point at which we wish to estimate $\varphi$. The dashed
line is the radial line from the origin of our spherical coordinate system
to P. The filled square on the surface is the interpolation point where we
evaluate $\rho(\theta_x, \phi_x)$.}
\label{fig:intp}
\end{figure}

\begin{figure}[h]
\begin{center}
  \epsffile{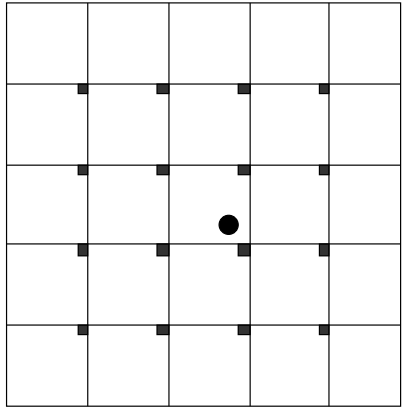}
\end{center}
\caption{This figure shows the choice of stencil points that we use for
biquartic interpolation for interpolation points that are on the interior of
the grid. The interpolation point is labelled by a filled circle and the
mesh points that are used as an interpolation stencil are denoted by filled
squares.}
\label{fig:neighbors1}
\end{figure}

\begin{figure}[h]
\begin{center}
  {\epsffile{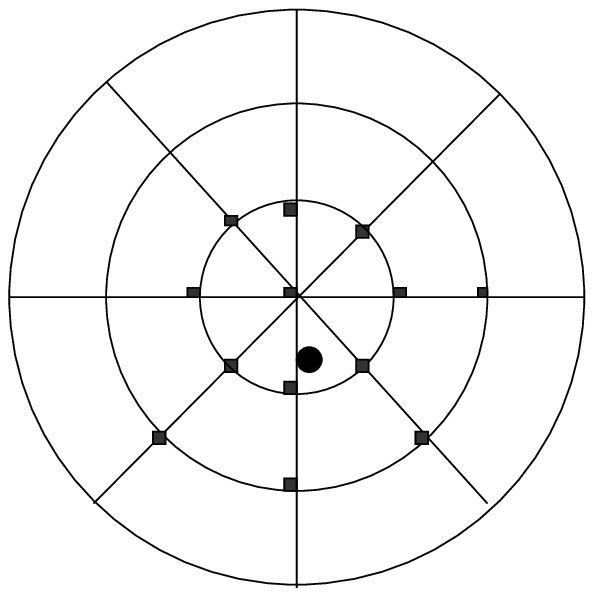}}
\end{center}
\caption{This figure shows the choice of stencil points for 
biquartic interpolation for interpolation points that are near the poles.
We view the pole points from a 3Dimensional perspective where the 
identifications of points in the $\phi$Direction is taken into account. This
leads to the special choice of interpolation stencil points as shown. This
leads to fourth order truncation error in the interpolant at the poles. 
The interpolation point is labelled by a filled circle and the
mesh points that are used as an interpolation stencil are denoted by filled
squares.}
\label{fig:neighbors2}
\end{figure}

To define $\varphi(x,y,z)$ at each stencil point ${\bf x \in {\cal N}}$ we 
use its split into radial and angular parts,
$\varphi(x,y,z) = r - \rho(\theta,\phi)$. For each stencil point ${\bf x}$ we 
compute
the corresponding spherical coordinates $(r_x, \theta_x, \phi_x)$. This point
can be thought of as a ray emanating from the origin of our spherical 
coordinate system (which coincides with the origin of our 
Cartesian coordinate system) along $(\theta_x, \phi_x)$ of length $r_x$. 
FIG.(\ref{fig:intp}) labels the point ${\bf x}$ as P. The dashed line from
P to the origin is the ray from the origin. Its intersection with 
$\hat{\cal S}$ is denoted by a filled square. The value of $\varphi$ at 
${\bf x}$ can be obtained by computing $\rho(\theta_x, \phi_x)$ via 
biquartic interpolation where the truncation error has a leading order term
which is fourth order in the grid spacing $h$. The interpolation is carried
out with values of $\hat{\rho}$ defined on mesh points
of $\hat{\cal S}$ using a 16 point stencil.  FIG.(\ref{fig:neighbors1}) 
shows the choice of these stencil points in the interior of the
mesh. At the poles a special choice is made of stencil points
which takes into account the indentifications made at the poles. 
FIG.(\ref{fig:neighbors2}) shows a choice of stencil points for an
interpolation point near the pole. This approach leads to a fourth order 
truncation error in $\rho(\theta_x,\phi_x)$ at all points on $\hat{\cal S}$.
Then $\varphi$ can be constructed for every ${\bf x \in {\cal N}}$ as
$\varphi = r_x - \rho(\theta_x, \phi_x)$. Using this approach $\varphi$ is
defined at any finite difference stencil point for every mesh point on 
$\hat{\cal S}$. The finite difference expressions for $\Delta^h_i \varphi$, 
$\Delta^h_i \Delta^h_j \varphi$ 
(corresponding to first and second derivatives) are 
computed at each of the mesh points.
The residual is then evaluated on $\hat{\cal S}$ using these finite
difference approximations for the derivatives to $O(h^2)$, and metric
data  ($\gamma_{ij}$, $\partial_k \gamma_{ij}$, $K_{ij}$) which are 
specified either 
analytically or interpolated from an enveloping 3D Cartesian grid.

Because we use $O(h^2)$ finite difference approximations $\Delta^h_i \varphi$,
$\Delta^h_i \Delta^h_j \varphi$ to the derivatives, this approach leads to an 
$O(h^2)$ truncation error in evaluating $\hat{F}[\hat{\rho}]$ . 
Because of our special attention to points near the pole, 
$\hat{F}$ is evaluated smoothly everywhere on $\hat{\cal S}$.

With a means for evaluating $\hat{F}$ at any point in the domain of $\hat{S}$
it is straightforward to generate $\hat{J}_{\bar{\mu}\bar{\nu}}$ numerically 
using Eq.(\ref{eqn:jacnumpert.1}). The algorithm for this is summarized as 
follows: 
{\sf
\begin{tabbing}
Specify metric data everywhere on $\hat{\cal S}$ \\
Evaluate $\hat{F}[\rho]$ everywhere on $\hat{\cal S}$ \\
For each point(labelled by $\bar{\nu})$ in $\hat{\cal S}$ \\
\hspace{1.0cm} \= Perturb $\rho_{\bar{\nu}} = \rho_{\bar{\nu}} + \epsilon$ \\
\> Specify metric data on perturbed point \\
\> Evaluate $\hat{F}[\rho_{\bar{\nu}} + \epsilon]$ at the $\bar{\mu}$-th
point. \\
\> Compute the $\bar{\mu}\bar{\nu}$ component of the Jacobian matrix \\
\>using (\ref{eqn:jacnumpert.1}) \\
foo\kill 
End loop over points on  $\hat{\cal S}$.
\end{tabbing}
}
This gives the Jacobian matrix, $\hat{J}_{\bar{\mu}\bar{\nu}}$, for $\hat{F}$ 
evaluated.
$\hat{J}_{\bar{\mu}\bar{\nu}}$ is a 
$(N_s^2 - 2N_s + 2)\times(N_s^2 - 2N_s + 2)$ matrix which is used in Newton's
method as follows:
{\sf
\begin{tabbing}
Start with an initial guess surface $\hat{\rho}=\hat{\rho}_0$ \\
while $\|\hat{F}\| >$ stopping criterion  \\
\hspace{1.0cm}\= Compute the Jacobian $\hat{J}_{\bar{\mu}\bar{\nu}}$ for the current $\hat{\rho}$ \\
\> Evaluate $\hat{F}[\hat{\rho}]$ \\
\> Solve $\hat{J}\cdot\delta \hat{\rho} = - \hat{F}[\hat{\rho}]$ 
for $\delta \hat{\rho}$ \\
\> Update the surface $\hat{\rho} = \hat{\rho} + \delta \hat{\rho}$\\
foo \kill
\end{tabbing}
}

\newpage
\onecolumn
\newpage
\newpage
\newpage
\newpage
\newpage
\newpage
\newpage
\newpage
\newpage
\newpage
\newpage
\newpage
\begin{figure}[h]
\caption{This figure shows plots  of the apparent horizon 
for the $a=0.0$ runs for $v=0.0,0.3,0.6$ and $0.9$ for boosts in the 
xyz-direction. The mesh used to find the apparent horizon is shown from
a top perspective. As the boost velocity is increased we see that the surface 
is contracted in the xyz-direction.}
\label{fig:kerrbst0_surf}
\end{figure}
\newpage
\newpage
\newpage
\newpage
\begin{figure}[h]
\caption{This figure shows plots  of the apparent horizon
for the $a=0.9$ runs for $v=0.0,0.3,0.6$ and $0.9$ for boosts in the
xyz-direction. The mesh used to find the apparent horizon is shown from
a top perspective. As the boost velocity is increased we see that the surface
is contracted in the xyz-direction.}
\label{fig:kerrbst9_surf}
\end{figure}
\newpage
\newpage
\newpage
\newpage
\newpage


\begin{thebibliography}{99}
\bibitem{York.frontiers}
J.W.~York in {\it Frontiers in Numerical Relativity}, ed. C.R.~Evans,
L.S.~Finn and D.W.~Hobill (Cambridge Univ. Press, 1989)
%
\bibitem{HawkingandEllis} Hawking, S. and Ellis, G.: ``The Large Scale
Structure of Space-Time'', Cambridge University Press, Cambridge (UK),
1973.
%
\bibitem{WaldandIyer} R. Wald and V. Iyer, Phys. Rev. D {\bf 44}, 
R3719-R3722 (1991).
%
\bibitem{nakamura.1} T. Nakamura, Y. Kojima, and K. Oohara, Phys. Lett.,
{\bf 106A}, 235-239, (1984).
%
%
\bibitem{bishop.1} N.T. Bishop, General Relativity and Gravitation,
{\bf 14}, No. 9, 1982.
%
%
\bibitem{anninos.1}
P.Anninos, K.Camarda, J.Libson, J.Masso,E.Seidel and W-M.Suen,
Phys. Rev. D {\bf 58} 024003 (1998).
%
\bibitem{baumgarte.1}
T. Baumgarte, G.B. Cook, M.A. Scheel, S.L. Shapiro, and S.A. Teukolsky,
``Implementing an Apparent Horizon Finder in Three-dimensions'',
Phys. Rev. D {\bf 54} 4849-4857 (1996).
%
\bibitem{thornburg.1} J. Thornburg, ``Finding apparent horizons in numerical
relativity'', Phys Rev D., {\bf 54}, 4899-4918 (1996).
%
\bibitem{Tod}
K.P. Tod, Class. Quantum Grav. {\bf 8}, L115-L118 (1991).
%
\bibitem{bernstein} D.H. Bernstein, ``Notes on the mean curvature flow method
for finding apparent horizons'', Unpublished notes.
%
\bibitem{pasch} E. Pasch, {\it SFB 382 } Report Number {\bf 63} (1997).
%
\bibitem{shoemaker} D.Shoemaker, M.F.Huq and R.A.Matzner,
{\it in preparation}.
%
\bibitem{Gundlach} C.Gundlach, ``Pseudospectral apparent horizon finders: An 
efficient new algorithm'', Phys Rev D {\bf 57} R863 (1998).
%
\bibitem{choptuik.4}
M.W. Choptuik, ``A Study of Numerical Techniques for Radiative Problems in
General Relativity'', Ph.D dissertation, The University of British Columbia,
July 1986.
%
%
\bibitem{Ortega} Ortega, J. and Rheinboldt, W., ``Iterative Solution of
Nonlinear Equations in Several Variables'', Academic Press, Inc., (USA)
 1970.
%
%
\bibitem{dongarra}
J.J. Dongarra, J.R. Bunch, C.B. Moler and G.W. Stewart,
``LINPACK : Users' Guide'', Philadelphia : Society for Industrial and Applied
Mathematics, 1978.
%

%
\bibitem{Thornburg.3}
Personal Communication
%

%
\bibitem{kershaw}
D. S. Kershaw, Journal of Comp. Phys., {\bf 26}, 43-65 (1978).
%
\bibitem{klasky_interpolators}
S.Klasky, M.Choptuik and R.Matzner, ``Interpolators/Extrapolators for AMR 
Finitie Difference codes'',
Technical Report, Center for Relativity, University of Texas at Austin (1995).
%
\bibitem{mhuq.texas.talk}
M.F.Huq, Third Texas Workshop on 3-dimensional Numerical Relativity,
The University of Texas at Austin, October 30th to November 1st 1995.

\bibitem{mhuq.phd}
M.F.Huq, ``Apparent Horizon Location in Numerical Spacetimes'',
Ph.D dissertation, The University of Texas at Austin, December 1996.

\bibitem{CFL.paper}
R. Courant, K.O. Friedrichs, and H. Lewy, English Translation of
"\"{U}ber die Partiellen Differenzengleichungen der Mathematischen Physik",
{\it Math. Ann.} 100, 32-74 (1928), IBM Journal, {\bf 11}(2) 215-238,
(March 1967).
%

%
\bibitem{MTW} C.W. Misner, K.S. Thorne and J.A. Wheeler, ``Gravitation'',
W.H. Freeman and Company, New York (1973).
%
\end{thebibliography}
\end{document}